\documentclass[11pt,a4paper]{article}

\usepackage[top=3.5cm,bottom=5.5cm,left=2.5cm,right=2.5cm]{geometry} % imposta i margini

\usepackage{array} % utile per gestire la larghezza delle colonne nelle tabelle

\usepackage{hyperref} % necessario per le referenze della bibliografia
\usepackage{amsmath,amssymb,amsthm,mathrsfs,textcomp,amscd,bbm} % necessario per molti comandi matematici

\usepackage{graphicx}
\DeclareGraphicsExtensions{.pdf,.png,.jpg,.mps,.ps,}
\graphicspath{{figures/}}

\usepackage{braket}

\newcommand{\be}{\begin{equation}}
\newcommand{\ee}{\end{equation}}
\newcommand{\bea}{\begin{eqnarray}}
\newcommand{\eea}{\end{eqnarray}}
\newcommand{\bi}{\begin{itemize}}
\newcommand{\ei}{\end{itemize}}

\newcommand{\bspl}{\begin{split}}
\newcommand{\espl}{\end{split}}

\newcommand{\MeV}{\,\mathrm{MeV}}

\newcommand{\GeV}{\,\mathrm{GeV}}
\newcommand{\fm}{\,\mathrm{fm}}

\def\mev{{\rm MeV}}
\def\gev{{\rm GeV}}
\def\tev{{\rm TeV}}
\def\fm{{\rm fm}}
 
\def\fm{\mathrm{fm}}
\def\ev{\mathrm{e\kern-0.1em V}}
\def\kev{\mathrm{ke\kern-0.1em V}}
\def\mev{\mathrm{Me\kern-0.1em V}}
\def\gev{\mathrm{Ge\kern-0.1em V}}
\def\tev{\mathrm{Te\kern-0.1em V}}

\begin{document}

\begin{center}

{\Large Tensor form factor of $D \to \pi(K) \ell \nu$ and $D \to \pi(K) \ell \ell$ decays\\[4mm] with $N_f=2+1+1$ twisted-mass fermions}

\vspace{1cm}

{\large V.~Lubicz$^{(a,b)}$, L.~Riggio$^{(b)}$, G.~Salerno$^{(a,b)}$, S.~Simula$^{(b)}$, C.~Tarantino$^{(a,b)}$}

\vspace{1cm}

$^{(a)}$ Dipartimento di Matematica e Fisica, Universit\'a di Roma Tre,\\ Via della Vasca Navale 84, I-00146 Roma, Italy\\[2mm]
$^{(b)}$ Istituto Nazionale di Fisica Nucleare, Sezione di Roma Tre,\\ Via della Vasca Navale 84, I-00146 Roma, Italy

\end{center}

\begin{figure}[htb!]
\centering{\includegraphics[scale=0.2]{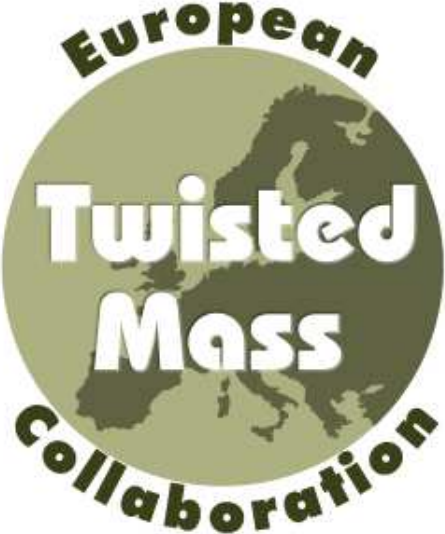}}
\end{figure}

\begin{abstract}
We present the first lattice $N_f = 2 + 1 + 1$ determination of the tensor form factor $f_T^{D \pi(K)}(q^2)$ corresponding to the semileptonic $D \to \pi(K) \ell \nu_\ell$ and rare $D \to \pi(K) \ell \ell$ decays as a function of the squared four-momentum transfer $q^2$.  
Together with our recent determination of the vector $f_+^{D \pi(K)}(q^2)$ and scalar  $f_0^{D \pi(K)}(q^2)$ form factors we complete the set of hadronic matrix elements regulating the semileptonic $D \to \pi(K) \ell \nu_\ell$ and rare $D \to \pi(K) \ell \ell$ transitions within and beyond the Standard Model, when a non-zero tensor coupling is possible. 
Our analysis is based on the gauge configurations produced by the European Twisted Mass Collaboration with $N_f = 2 + 1 + 1$ flavors of dynamical quarks, which include in the sea, besides two light mass-degenerate quarks, also the strange and charm quarks with masses close to their physical values. 
We simulated at three different values of the lattice spacing and with pion masses as small as $220$ MeV and with the valence heavy quark in the mass range from $\simeq 0.7\, m_c^{phys}$ to $\simeq 1.2\, m_c^{phys}$. 
The matrix elements of the tensor current are determined for a plethora of kinematical conditions in which parent and child mesons are either moving or at rest. 
As in the case of the vector and scalar form factors, Lorentz symmetry breaking due to hypercubic effects is clearly observed also in the data for the tensor form factor and included in the decomposition of the current matrix elements in terms of additional form factors. 
After the extrapolations to the physical pion mass and to the continuum and infinite volume limits we determine the tensor form factor in the whole kinematical region from $q^2 = 0$ up to $q^2_{\rm max} = (M_D - M_{\pi(K)})^2$ accessible in the experiments.
A set of synthetic data points, representing our results for $f_T^{D \pi(K)}(q^2)$ for several selected values of $q^2$, is provided and the corresponding covariance matrix is also available.
At zero four-momentum transfer we get $f_T^{D \pi}(0) = 0.506 ~ (79)$ and $f_T^{D K}(0) = 0.687 ~ (54)$, which correspond to $f_T^{D \pi}(0) / f_+^{D \pi}(0) = 0.827 ~ (114)$ and $f_T^{D K}(0) / f_+^{D K}(0)= 0.898 ~ (50)$. 
\end{abstract}

\clearpage

%%%%%%%%%%%%%%%%%%%%%%%%%%%%%%%
\section{Introduction}
\label{sec:intro}
%%%%%%%%%%%%%%%%%%%%%%%%%%%%%%%

Precise measurements of hadron weak decays can constrain the Standard Model (SM) and place bounds on New Physics (NP) models.
The semileptonic and rare transitions between pseudoscalar (P) mesons can be parametrized, in all extensions of the SM, in terms of three form factors, namely the vector $f_+$, the scalar $f_0$ and the tensor $f_T$ ones. 
New particles beyond the SM, such as those appearing in NP models with supersymmetry or in a fourth generation or in composite Higgs sectors, can alter the Wilson coefficients of the effective weak Hamiltonian that describes physics below the electroweak scale.
Whatever these unknown particles may be, the hadronic physics remains the same.

Recently in Ref.~\cite{Lubicz:2017syv} we presented the first $N_f = 2 + 1 + 1$ lattice QCD (LQCD) calculation of the vector and scalar form factors $f_+^{D \pi(K)}(q^2)$ and $f_0^{D \pi(K)}(q^2)$ governing the semileptonic $D \to \pi(K) \ell \nu$ decays within the SM.
We employed the gauge configurations generated by the European Twisted Mass Collaboration (ETMC) with $N_f = 2 + 1 + 1$ dynamical quarks, which include in the sea, besides two light mass-degenerate quarks, also the strange and charm quarks with masses close to their physical values \cite{Baron:2010bv,Baron:2011sf}. 
In this work we complete the set of operators relevant for the $D \to \pi(K)$ transitions by analyzing the matrix elements of the weak tensor current $\bar{c} \sigma_{\mu\nu} d(s)$ to obtain the tensor form factor $f_T^{D \pi(K)}(q^2)$. 
The latter may enter both in the semileptonic $D \to \pi(K) \ell \nu$ decays as a contribution beyond the SM and in the rare decays driven by the transition $c \to u \ell^+ \ell^-$, which are loop-suppressed in the SM as they proceed through flavor-changing neutral currents.

Following Ref.~\cite{Lubicz:2017syv} we have evaluated the tensor form factor $f_T^{D \pi(K)}(q^2)$ in the whole range of values of $q^2$ accessible in the experiments, i.e.~from $q^2 = 0$ up to $q^2_{\rm max} = (M_D - M_{\pi(K)})^2$.
In our calculations quark momenta are injected on the lattice using non-periodic boundary conditions \cite{Bedaque:2004kc,deDivitiis:2004kq} and the matrix elements of the tensor current are determined for plenty of kinematical conditions, in which parent and child mesons are either moving or at rest.

The data coming from different kinematical conditions exhibit a remarkable breaking of Lorentz symmetry due to hypercubic effects for both $D\to \pi$ and $D\to K$ form factors. 
The presence of these effects was already observed in Ref.~\cite{Lubicz:2017syv} in the case of the vector and scalar form factors, and there we presented a method to subtract the hypercubic artifacts and to recover the Lorentz-invariant form factors in the continuum limit.

Besides Ref.~\cite{Lubicz:2017syv} hypercubic effects were never observed in the context of the $D \to \pi(K)$ transitions. 
Previous lattice calculations used only a limited number of kinematical conditions (typically the $D$-meson at rest). 
This limitation obscures the presence of hypercubic effects in the lattice data. 
Moreover, in Ref.~\cite{Lubicz:2017syv} we found that the hypercubic artifacts strongly depend on the difference between the parent and the child meson masses. 
This is an important issue, which warrants dedicated investigations. 
If this is the case, the hypercubic artifacts may play an important role in the determination of the form factors governing the semileptonic $B$-meson decays and it is therefore crucial to have them under control.

In Ref.~\cite{Lubicz:2017syv} the subtraction of the hypercubic effects was achieved by considering a decomposition of the current matrix element which contains, beside the usual Lorentz-covariant part, additional hypercubic structures proportional to $a^2$. 
The form of this decomposition depends on the Dirac structure of the current and, therefore, it is interesting to further validate the method by applying it also to the case of the tensor current.

In this work we present the subtraction of the hypercubic artifacts and the determination of the Lorentz-invariant tensor form factor $f_T^{D \pi(K)}(q^2)$ after the combined extrapolations to the physical pion mass and to the continuum limit.
A set of synthetic data points, representing our results for $f_T^{D \pi(K)}(q^2)$ for several selected values of $q^2$, is provided (see later on Tables \ref{tab:synthetic_Dpi} and \ref{tab:synthetic_DK}).
The full covariance matrix, which includes the synthetic data points of the vector and scalar form factors obtained in Ref.~\cite{Lubicz:2017syv} and those of the tensor form factor from the present work, is available upon request.

At zero four-momentum transfer our results are
 \be
    \label{eq:fT0}
    f_T^{D\pi}(0) = 0.506 ~ (79) ~ , \qquad \qquad  f_T^{DK}(0) = 0.687 ~ (54)
 \ee
and
  \be
    \label{eq:fT0f+0}
    \frac{f_T^{D\pi}(0)}{f_+^{D\pi}(0)} = 0.827 ~ (114) ~ ,  \qquad \qquad \frac{f_T^{DK}(0)}{f_+^{DK}(0)} = 0.898 ~ (50) ~ ,
 \ee
where the errors include both statistical and systematic uncertainties, added in quadrature.

The paper is organized as follows. 
In Section~\ref{sec:simulations} we describe the simulation details.
In Section~\ref{sec:sec1} we present the computation of the tensor form factors $f_T^{D \pi(K)}(q^2)$  using the matrix elements of the weak tensor current relevant for the $D \to \pi(K)$ transition, obtained from two-point and three-point correlation functions. 
In Section~\ref{sec:sec2} the evidence of Lorentz symmetry breaking in the momentum dependence of the form factors is presented and discussed. 
In Section~\ref{sec:sec3} we describe the strategy adopted in order to extract the physical, Lorentz invariant, tensor form factors. 
This is based on a global fit of the data corresponding to all lattice ensembles, studying simultaneously the dependence on $q^2$, the light-quark mass $m_\ell$ and the lattice spacing $a$, and using a phenomenological Ansatz to describe the hypercubic effects.
In Section~\ref{sec:sec4} the results for $f_T^{D \pi(K)}(q^2)$ from the global fit, as well as for the ratio $f_T^{D \pi(K)}(q^2) / f_+^{D \pi(K)}(q^2)$ are shown.
Finally our  conclusions are summarized in Sec.~\ref{sec:conclusions}.

%%%%%%%%%%%%%%%%%%%%%%%%%%%%%%%
\section{Simulation details}
\label{sec:simulations}
%%%%%%%%%%%%%%%%%%%%%%%%%%%%%%%

The gauge ensembles used in this work have been generated by the ETMC with $N_f = 2 + 1 + 1$ dynamical quarks, which include in the sea, besides two light mass-degenerate quarks, also the strange and the charm quarks \cite{Baron:2010bv,Baron:2011sf}.
The ensembles and the simulations are the same adopted in Ref.~\cite{Lubicz:2017syv} for the determination of the vector $f_+^{D\pi(K)}(q^2)$ and scalar $f_0^{D\pi(K)}(q^2)$ form factors.
Here, in Table~\ref{tab:simu&masses} we recall the basic simulation parameters and the masses of the  $\pi$, $K$ and $D$ mesons corresponding to each ensemble.

\begin{table}[!htb]
\renewcommand{\arraystretch}{1.2}	 
\begin{center}	
\scalebox{0.85}{
\begin{tabular}{||>{$\!\!$}c<{$\!\!$}|>{$\!\!$}c<{$\!\!$}|>{$\!\!$}c<{$\!\!$}|>{$\!\!$}c<{$\!\!$}|>{$\!\!$}c<{$\!\!$}|>{$\!\!$}c<{$\!\!$}|>{$\!\!$}c<{$\!\!$}|>{$\!\!$}c<{$\!\!$}|>{$\!\!$}c<{$\!\!$}|>{$\!\!$}c<{$\!\!$}|>{$\!\!$}c<{$\!\!$}||}
\hline
ensemble&$\beta$ & $V / a^4$ &$a\mu_{sea}=a\mu_\ell$ &$a\mu_s$&$a\mu_c$&$M_\pi(\MeV)$&$M_K(\MeV)$& $M_D(\MeV)$& $L(\fm)$&$M_\pi L$\\
\hline
A30.32&$1.90$ & $32^{3}\times 64$ &$0.0030$ &$\{0.0180,$&$\{0.21256,$&$275$&$569$&$2015$&$2.84$&$3.96$\\
A40.32&& & $0.0040$ &$0.0220,$ &$0.25000,$&$315$&$578$&$2018$&$$&$4.53$\\
A50.32&& & $0.0050$ &$ 0.0260\}$ &$0.29404\}$&$351$&$578$&$2018$&$$&$5.04$\\
\cline{1-1}\cline{3-4}\cline{7-11}
A40.24&& $24^{3}\times 48 $ & $0.0040$ & & &$324$&$584$&$2024$&$2.13$&$3.49$\\
A60.24&& & $0.0060$ && &$386$&$599$&$2022$&$$&$4.17$\\
A80.24&& & $0.0080$ & &  &$444$&$619$&$2037$&$$&$4.79$\\
A100.24&& & $0.0100$ & &  &$495$&$639$&$2042$&$$&$5.34$\\
\hline
B25.32&$1.95$ & $32^{3}\times 64$ &$0.0025$ & $\{0.0155,$& $\{0.18705,$&$258$&$545$&$1950$&$2.61$&$3.42$\\
B35.32&& & $0.0035$ &$  0.0190,$ &$0.22000,$&$302$&$556$&$1944$&$$&$3.99$\\
B55.32&& & $0.0055$ &$ 0.0225\}$ &$0.25875\}$&$375$&$578$&$1959$&$$&$4.96$\\
B75.32&& & $0.0075$ & & &$436$&$600$&$1965$&$$&$5.77$\\
\cline{1-1}\cline{3-4}\cline{7-11}
B85.24&& $24^{3}\times 48 $ & $0.0085$  & & &$467$&$611$&$1974$&$1.96$&$4.63$\\
\hline
D15.48&$2.10$ & $48^{3}\times 96$ &$0.0015$ & $\{0.0123,$& $\{0.14454,$ &$220$&$526$&$1928$&$2.97$&$3.31$\\
D20.48&& & $0.0020$ &$0.0150,$ &$0.17000,$&$254$&$533$&$1933$&$$&$3.83$\\
D30.48&& & $0.0030$ & $  0.0177\}$&$0.19995\}$&$308$&$547$&$1939$&$$&$4.65$\\
\hline
\end{tabular}
}
\end{center}
\renewcommand{\arraystretch}{1.0}
\vspace{-0.25cm}
\caption{\it \small Summary of the simulated sea and valence quark bare masses, of the $\pi$, $K$ and $D$ meson masses, of the lattice size $L$ and of the product $M_\pi L$ for the various gauge ensembles used in this work. The values of $M_K$ and $M_D$ do not correspond to the simulated strange and charm bare quark masses shown in the $5^{\rm th}$ and $6^{\rm th}$ columns, but to the renormalized strange and charm masses interpolated at the physical values $m_s^{phys}(\overline{MS},2~\rm{GeV}) = 99.6 (4.3) \MeV$ and $m_c^{phys}(\overline{MS}, 2~\rm{GeV}) = 1.176 (39) \GeV$ determined in Ref.~\cite{Carrasco:2014cwa}. After Ref.~\cite{Lubicz:2017syv}.}
\label{tab:simu&masses}
\end{table} 

The gauge fields are simulated using the Iwasaki gluon action \cite{Iwasaki:1985we}, while sea quarks are implemented with the Wilson Twisted Mass Action at maximal twist \cite{Frezzotti:2000nk,Frezzotti:2003xj,Frezzotti:2003ni}. 
In order to avoid the mixing of strange and charm quarks induced by lattice 
artifacts in the unitary twisted-mass formulation we have adopted the non-unitary setup described in Ref.~\cite{Frezzotti:2004wz}, in which the valence strange quarks are regularized as Osterwalder-Seiler (OS) fermions \cite{Osterwalder:1977pc}, while the valence up and down quarks have the same action as the sea. 
The use of different lattice regularisations for the valence and sea quarks of the second generation preserves unitarity in the continuum limit and does not modify the operator renormalization pattern in mass-independent schemes, while producing only a modification of discretization effects.
Moreover, since we work at maximal twist, physical observables are guaranteed  to be automatically ${\cal{O}}(a)$-improved~\cite{Frezzotti:2003ni,Frezzotti:2004wz}.

The QCD simulations have been carried out at three different values of the inverse bare lattice coupling $\beta$, to allow for a controlled extrapolation to the continuum limit, and at different lattice volumes.  
We have simulated quark masses in the range from $\simeq 3\, m_{ud}^{phys}$ to $\simeq 12\, m_{ud}^{phys}$ in the light sector, from $\simeq 0.7\, m_s^{phys}$ to $\simeq 1.2\, m_s^{phys}$ in the strange sector, and from $\simeq 0.7\, m_c^{phys}$ to $\simeq 1.2\, m_c^{phys}$ in the charm sector, where $m_{ud}^{phys}$, $m_s^{phys}$ and $m_c^{phys}$ are the physical values of the average up/down, strange and charm quark masses respectively, as determined in Ref.~\cite{Carrasco:2014cwa}.
The lattice scale is fixed using as input the experimental value of the pion decay constant $f_\pi$ from PDG~\cite{Olive:2016xmw}. 
The values of the lattice spacing are: $a = \{ 0.0885\,(36), 0.0815\,(30), 0.0619\,(18)\}\, \fm$ at $\beta = \{1.90, 1.95, 2.10\}$ respectively.
The lattice volume goes from $\simeq 2$ to $\simeq 3$ fm and the pion masses range from $\simeq 220$ to $ \simeq 500\,\MeV $.

Following Ref.~\cite{Lubicz:2017syv} we make use of the input parameters (values of quark masses and lattice spacings) obtained from the eight branches of the analysis carried out in Ref.~\cite{Carrasco:2014cwa}. 
The various branches differ by: 
~ i) the choice of the scaling variable, which was taken to be either the Sommer parameter $r_0/a$ \cite{Sommer:1993ce} or the mass of a fictitious P-meson made of two valence strange-like quarks $a M_{s^\prime s^\prime}$; 
~ ii) the fitting procedures, which were based either on Chiral Perturbation Theory (ChPT) or on a polynomial expansion in the light quark mass (for the motivations see the discussion in Section 3.1 of Ref.~\cite{Carrasco:2014cwa}); 
~ iii)  the choice between two methods, denoted as M1 and M2 which differ by ${\cal{O}}(a^2)$ effects (see, e.g., Ref.~\cite{Constantinou:2010gr}), used to determine non-perturbatively in the RI$^\prime$-MOM scheme the values of the mass renormalization constant (RC) ${\cal{Z}}_m = 1 / {\cal{Z}}_P$~\cite{Carrasco:2014cwa}. The use of the two sets of RCs lead to the same final results once the continuum limit for the physical quantity of interest is performed. 

In what follows we make use of the local version of the tensor operator, which in the twisted-mass setup renormalizes multiplicatively and is ${\cal{O}}(a)$-improved~\cite{Frezzotti:2003ni}, namely
 \be
     \widehat{T}_{\mu \nu} \equiv {\cal{Z}}_T \bar{c} \sigma_{\mu \nu} q ~ ,
     \label{eq:tensor_local}
 \ee
where $q = \{d, s\}$.
The tensor current RC, ${\cal{Z}}_T$, was computed in the RI$^\prime$-MOM scheme and converted in the $\overline{MS}$ scheme at a renormalization scale equal to $2$ GeV in Ref.~\cite{Carrasco:2014cwa}.
The numerical values of ${\cal{Z}}_T$ used in our analyses are listed in Table~\ref{tab:Z_T}.

\begin{table}[!htb]
\renewcommand{\arraystretch}{1.2} 
\begin{center}
\begin{tabular}{||c|c|c||}
\hline
$\beta$ & ${\cal{Z}}_T^{\overline{MS}}(2\,\mathrm{GeV})$ & ${\cal{Z}}_T^{\overline{MS}}(2\,\mathrm{GeV})$ \\ 
             & M1 method & M2 method \\ \hline \hline
$1.90$ & $0.711(5)$ & $0.700(3)$ \\ \hline
$1.95$ & $0.724(4)$ & $0.711(2)$ \\ \hline
$2.10$ & $0.774(4)$ & $0.767(2)$ \\ \hline
\end{tabular}
\end{center}
\renewcommand{\arraystretch}{1.0}
\vspace{-0.25cm}
\caption{\it \small Input values for the tensor current RC ${\cal{Z}}_T^{\overline{\rm MS}}(2~\gev)$ corresponding to the methods M1 and M2, computed in Ref.~\cite{Carrasco:2014cwa}.}
\label{tab:Z_T}
\end{table} 

Throughout this work the results corresponding to the eight branches of the analysis are combined to form our averages and errors according to Eq.~(28) of Ref.~\cite{Carrasco:2014cwa}.

%%%%%%%%%%%%%%%%%%%%%%%%%%%%%%%
\section{Lattice calculation of the tensor matrix elements}
\label{sec:sec1}
%%%%%%%%%%%%%%%%%%%%%%%%%%%%%%%

The matrix element of the renormalized tensor current (\ref{eq:tensor_local}) between an initial $D$-meson state and a $\pi$($K$)-meson final state can be written, as required by the Lorentz symmetry, in terms of a single form factor $f_T^{D\pi(K)}(q^2)$:
\be
    \langle P(p_P) | \widehat{T}^{\mu \nu} | D(p_D) \rangle  = \frac{2}{M_D + M_P} 
        \left[ p_P^\mu p_D^\nu  - p_P^\nu p_D^\mu \right] f_T^{DP}(q^2) ~ ,
    \label{eq:matrix_element_decomposition}
\ee
where $P = \pi (K)$ can be either the pion or the kaon and the four-momentum transfer $q$ is given by $q \equiv p_D - p_P$.
The factor $2 / (M_D + M_P)$ is conventionally inserted to make the tensor form factor dimensionless. 

In order to inject momenta on the lattice we use the same procedure adopted in Refs.~\cite{Lubicz:2017syv,Carrasco:2016kpy} for the $D \to \pi(K)$ semileptonic decays and the $K_{\ell 3}$ decays. 
In particular for the valence quark fields we impose twisted boundary conditions (BC's) \cite{Bedaque:2004kc,deDivitiis:2004kq,Guadagnoli:2005be} in the spatial directions in order to remove the limitations resulting from the use of periodic BC's and to access the whole kinematical region for momentum-dependent quantities, like the form factors. 
The sea quarks, on the contrary, have been simulated in Refs.~\cite{Baron:2010bv,Baron:2011sf} with periodic BC's in space. 
As shown in Refs.~\cite{Sachrajda:2004mi,Bedaque:2004ax}, for physical quantities which do not involve final state interactions (like, e.g., meson masses, decay constants and semileptonic form factors), the use of different BC's for valence and sea quarks produces only finite size effects (FSEs) which are exponentially small.
For both valence and sea quarks the BC's in time are anti-periodic.
Thus, the valence quark three-momentum is given by
 \be
    p = \frac{2 \pi}{L}\left( \theta + n_x , ~ \theta + n_y , ~ \theta + n_z \right) ~ ,
    \label{eq:pj_twisted}
  \ee
where $n_{x,y,z}$ are integers and the $\theta$ values are equal in all the three spatial directions, listed in Table~\ref{tab:theta_values} for the reader's convenience. 
\begin{table}[!htb] 
\renewcommand{\arraystretch}{1.2} 
\begin{center}
\begin{tabular}{||c|c||c||}
\hline
$\beta$ & $V / a^4$ & $\theta$\\
\hline
$1.90$ & $32^{3} \times 64$ & $0.0, ~ \pm 0.200, ~ \pm0 .467, ~ \pm 0.867$\\
\cline{2-3}
            & $24^{3} \times 48 $ & $0.0, ~ \pm 0.150, ~ \pm 0.350, ~ \pm 0.650$\\
\hline
$1.95$ & $32^{3} \times 64$ & $0.0, ~ \pm 0.183, ~ \pm 0.427, ~ \pm 0.794$\\
\cline{2-3}
            & $24^{3} \times 48 $ &$0.0, ~ \pm 0.138, ~ \pm 0.321, ~ \pm 0.596$\\
\hline
$2.10$ & $48^{3} \times 96$ &$0.0, ~ \pm 0.212, ~ \pm 0.493, ~ \pm 0.916$\\
\hline
\end{tabular} 
\end{center}
\renewcommand{\arraystretch}{1.0} 
\vspace{-0.25cm}
\caption{\it \small Values of the parameter $\theta$, appearing in Eq.~(\ref{eq:pj_twisted}), for the various ETMC gauge ensembles (after Ref.~\cite{Lubicz:2017syv}).}
\label{tab:theta_values}
\end{table} 
 
The matrix elements $\langle P(p_P) | \widehat{T}_{\mu \nu} | D(p_D) \rangle$ can be extracted from the large (Euclidean) time distance behavior of a convenient combination of two-point and three-point lattice correlation functions, defined as 
 \bea
   \label{eq:C2}
   C_2^{D(P)}(t^\prime, \vec{p}_{D(P)}) &=& \frac{1}{L^3} \sum_{\vec{x},\vec{z}} 
         \braket{0 \lvert P_5^{D(P)}(x) P_5^{D(P)\dagger}(z) \rvert 0}
         e^{-i\vec{p}_{D(P)}\cdot(\vec{x} - \vec{z})} \delta_{t^\prime, t_x - t_z} , ~ \\[2mm]
   \label{eq:C3}
   C_{\widehat{T}_{\mu \nu}}^{DP}(t, t^\prime, \vec{p}_D, \vec{p}_P) &=& \frac{1}{L^6} 
        \sum_{\vec{x}, \vec{y},\vec{z}} \braket{0 \lvert P_5^P(x) \widehat{T}_{\mu \nu}(y) 
        P_5^{D\dagger}(z) \rvert 0} e^{-i\vec{p}_{D} \cdot (\vec{y} - \vec{z}) + 
        i \vec{p}_{P} \cdot (\vec{y} - \vec{x})} \delta_{t^\prime, t_x - t_z} \delta_{t, t_y - t_z} , \quad
 \eea
where $t^\prime$ is the time distance between the sink and the source, $t$ is the time distance between the insertion of the tensor current $\widehat{T}_{\mu \nu}$ and the source, and $P_5^D = i \bar{c} \gamma_5 u$ and $P_5^{\pi(K)} = i \bar{d}(\bar{s}) \gamma_5 u$ are the interpolating fields of the $D$ and $\pi(K)$ mesons. 
The Wilson parameters of the two valence quarks in the parent and child mesons are always chosen to have opposite values, i.e.~$r_c = r_s = r_d = -r_u$.
In this way the squared $\pi(K)$-meson mass differs from its continuum counterpart only by terms of order ${\cal{O}}(a^2 \mu_{\ell(s)} \Lambda_{QCD})$ \cite{Frezzotti:2003ni}.

The three-point correlation functions $C_{\widehat{T}_{\mu \nu}}^{DP}(t, t^\prime, \vec{p}_D, \vec{p}_P)$ have been simulated by imposing periodic BC's to the spectator valence $u$ quark and twisted BC's (\ref{eq:pj_twisted}) to both the initial $c$ and final $d(s)$ quarks. 
With this choice the $\pi$, $K$ and $D$ meson (spatial) three-momenta are given by 
 \be
      \vec{p}_{D(P)} = \frac{2 \pi}{L} \left( \theta_{D(P)}, \theta_{D(P)}, \theta_{D(P)} \right)
      \label{eq:pmeson_twisted}
 \ee      
where $\theta_{D(P)}$ can assume for each gauge ensemble the values of the parameter $\theta$ given in Table~\ref{tab:theta_values}, which have been chosen in order to obtain values of $|\vec{p}_{D(P)}|$ equal to $\approx 150\MeV$, $\approx 350\MeV$ and $\approx 650\MeV$ independently of the lattice spacing and volumes.

Since we have used only momenta distributed democratically in the three spatial directions, the only non-vanishing matrix elements of the tensor current are $\langle P(p_P) | \widehat{T}^{0 j} | D(p_D) \rangle$ with $j = 1, 2, 3$, which are equal to each other. 
Therefore, in order to improve the statistics, in what follows we introduce the tensor operator $\widehat{T}_{sp}$ given by
 \be
       \label{eq:Tsp}
      \widehat{T}_{sp} \equiv \frac{1}{3} \sum_{j=1}^3 \widehat{T}^{0 j} ~ .
 \ee

The statistical accuracy of the correlators (\ref{eq:C2}-\ref{eq:C3}) is significantly improved by using the all-to-all quark propagators evaluated with the so-called ``one-end" stochastic method \cite{McNeile:2006bz}, which includes spatial stochastic sources at a single time slice chosen randomly (see Ref.~\cite{Frezzotti:2008dr} where the degenerate case of the electromagnetic pion form factor is discussed in details).
Statistical errors on the quantities directly extracted from the correlators are always evaluated using the jackknife procedure, while cross-correlations are taken into account by the use of the eight bootstrap samplings (with ${\cal{O}}(100)$ events each) corresponding to the eight analyses of Ref.~\cite{Carrasco:2014cwa} (see Section~\ref{sec:simulations}).

In the case of charm quarks we adopt Gaussian-smeared interpolating fields~\cite{Gusken:1989qx} for both the source and the sink in order to suppress faster the contribution of the excited states, leading to an improved projection onto the ground state at relatively small time distances. 
For the values of the smearing parameters we set $k_G = 4$ and $N_G = 30$. 
In addition, we apply APE-smearing to the gauge links~\cite{Albanese:1987ds} in the interpolating fields with parameters $\alpha_{APE} = 0.5$ and $N_{APE} = 20$.
The Gaussian smearing is applied as well also for the light and strange quarks. 

As is well known, at large time distances the two-point and three-point correlation functions behave as
 \bea
        \label{eq:C2_larget}
        C_2^{D(P)}(t^\prime, \vec{p}_{D(P)}) & ~ _{\overrightarrow{t^\prime \gg a}} ~ & 
            \frac{|Z_{D(P)}|^2}{2E_{D(P)}} \left[ e^{-E_{D(P)} t^\prime} + e^{-E_{D(P)} (T - t^\prime)} \right] , \\
        \label{eq:C3_larget}        
        C_{\widehat{T}_{sp}}^{DP}(t, t^\prime, \vec{p}_D, \vec{p}_P) 
            & ~ _{\overrightarrow{t\gg a, (t^\prime-t)\gg a}} ~ & \frac{Z_P Z_D^*}{4E_P E_D} 
            \braket{P(p_P) |\widehat{T}_{sp}|D(p_D)} e^{-E_D t} e^{-E_P (t^\prime - t)} , \quad
 \eea
where $E_{D}$ and $E_{P}$ are the energies of the $D$ and $P$ mesons, while $Z_D$ and $Z_P$ are the matrix elements $\braket{0\lvert\,P_5^D(0)\,\rvert\,D(\vec{p}_D) }$ and $\braket{0\lvert\,P_5^P(0)\,\rvert\,P(\vec{p}_{P})}$, which depend on the meson momenta $\vec{p}_D$ and $\vec{p}_P$ because of the use of smeared interpolating fields. 
The matrix elements $Z_D$ and $Z_P$ can be extracted directly by fitting the large time behavior of the corresponding two-point correlation functions using the exponential behavior given by the r.h.s.~of Eq.~(\ref{eq:C2_larget}). 
The time intervals $[t_{\rm min}, t_{\rm max}]$ adopted for the fit (\ref{eq:C2_larget}) are taken consistently with the one used in Ref.~\cite{Lubicz:2017syv} for the calculation of the vector and scalar form factors $f_+^{D \pi(K)}(q^2)$ and $f_0^{D \pi(K)}(q^2)$, and they are listed in Table~\ref{tab:time_intervals}.
\begin{table}[!htb]
\renewcommand{\arraystretch}{1.2} 
\begin{center}
\begin{tabular}{||c|c|c|c||c||}
\hline
$\beta$ & $V / a^4$ &$[t_{\rm min},\,t_{\rm max}]_{(\ell\ell,\,\ell s)}/a$&$[t_{\rm min},\,t_{\rm max}]_{(\ell c)}/a$&$t^\prime/a$\\
\hline
$1.90$ & $32^{3}\times 64$ &$[12,\,31]$&$[8,\,16]$&$18$ \\
& $24^{3}\times 48 $ & $[12,\,23]$&$[8,\,17]$&$18$\\
\hline
$1.95$ & $32^{3}\times 64$ & $[13,\,31]$&$[9,\,18]$&$20$\\
& $24^{3}\times 48 $ &$[13,\,23]$&$[9,\,18]$&$20$\\
\hline
$2.10$ & $48^{3}\times 96$ &$[18,\,40]$&$[12,\,24]$&$26$\\
\hline
\end{tabular}
\end{center}
\renewcommand{\arraystretch}{1.0}
\vspace{-0.25cm}
\caption{\it \small Time intervals adopted for the extraction of the P meson energies $E_{D(P)}$ and the matrix elements $Z_{D(P)}$ from the two-point correlators in the light ($\ell$), strange ($s$) and charm ($c$) sectors. The last column contains the values of the time distance $t^\prime$ between the source and the sink adopted for the three-point correlators (\ref{eq:C3}).}
 \label{tab:time_intervals}
\end{table} 

The energies $E_{D(P)}$ extracted from the fit (\ref{eq:C2_larget}) are consistent (within the statistical errors) with the continuum-like dispersion relation $E_{D(P)}^{\rm disp} = \sqrt{M_{D(P)}^2 + |\vec{p}_{D(P)}|^2}$, where $M_{D(P)}$ is the meson mass extracted from the two-point correlator corresponding to the meson at rest.
%The differences are within one standard deviation in the case of $E_\pi$ and $E_K$, and $\simeq 2$ standard deviations for $E_D$.
As in Ref.~\cite{Lubicz:2017syv}, we use for the analysis the energy values $E_{D(P)}^{\rm disp}$ instead of those directly extracted from the fit.

We have evaluated the three-point correlators (\ref{eq:C3}) for various choices of the time distance $t^\prime$ between the source and the sink in the case of a representative subset of the ETMC gauge ensembles.
In this way the choice of $t^\prime$ has been optimized in order to reduce the statistical noise keeping at the same time that the ground-state signals (\ref{eq:C3_larget}) change only within the statistical errors. 
The values adopted for $t^\prime$ are found to be the same as those used for the vector and scalar form factors in Ref.~\cite{Lubicz:2017syv} and are shown in the last column of Table~\ref{tab:time_intervals}.

The matrix elements $\braket{P(p_P) |\widehat{T}_{sp}|D(p_D)}$ can be extracted from the time dependence of the ratio $R$ between the two-point and three-point correlation functions given in Eqs.~(\ref{eq:C2_larget}-\ref{eq:C3_larget}), namely
 \be
    \label{eq:R} 
    R(t, \vec{p}_D,\vec{p}_P) \equiv 4 E_D E_P \frac{C_{T_{sp}}^{DP}(t, t^\prime, \vec{p}_D,\vec{p}_P)
        ~ C_{T_{sp}}^{PD}(t, t^\prime, \vec{p}_P,\vec{p}_D)}{\widetilde{C}_2^D(t^\prime, \vec {p}_D) ~ 
        \widetilde{C}_2^P(t^\prime, \vec{p}_P)} ~ ,
 \ee
where the correlation function $\widetilde{C}_2^{D(P)}$ is given by
 \be
      \widetilde{C}_2^{D(P)}(t, \vec{p}_{D(P)}) \equiv \frac{1}{2}\left[ C_2^{D(P)}(t, \vec{p}_{D(P)}) + 
          \sqrt{C_2^{D(P)}(t, \vec{p}_{D(P)})^2 - C_2^{D(P)}(\frac{T}{2}, \vec{p}_{D(P)})^2} \right] ~ ,
      \label{eq:C2_tilde}
 \ee
which at large time distances behave as
 \be
      \widetilde{C}_2^{D(P)}(t, \vec{p}_{D(P)}) ~ _{\overrightarrow{t \gg a}} ~ \frac{Z_{D(P)}}{2 E_{D(P)}}
          e^{-E_{D(P)} t} ~ ,
     \label{eq:C2_tilde_larget}
 \ee
i.e.~without the backward signal.
One has:
\be
    \label{eq:R_plateau}
    R(t,\vec{p}_D, \vec{p}_P) _{\overrightarrow{t\gg a, (t^\prime-t)\gg a}} ~ \lvert\braket{P(p_P)| \widehat{T}_{sp} |D(p_D)}\rvert^2 ~ .
\ee
Taking the  square root of $R$ we can get the absolute value of the matrix elements $\braket{P(p_P) |\widehat{T}_{sp}|D(p_D)}$, while its sign can be easily inferred from that of the correlator $C_{T_{sp}}^{DP}(t, t^\prime, \vec{p}_D, \vec{p}_P)$ in the relevant time regions.

\begin{figure}[htb!]
\centering
\includegraphics[scale=0.60]{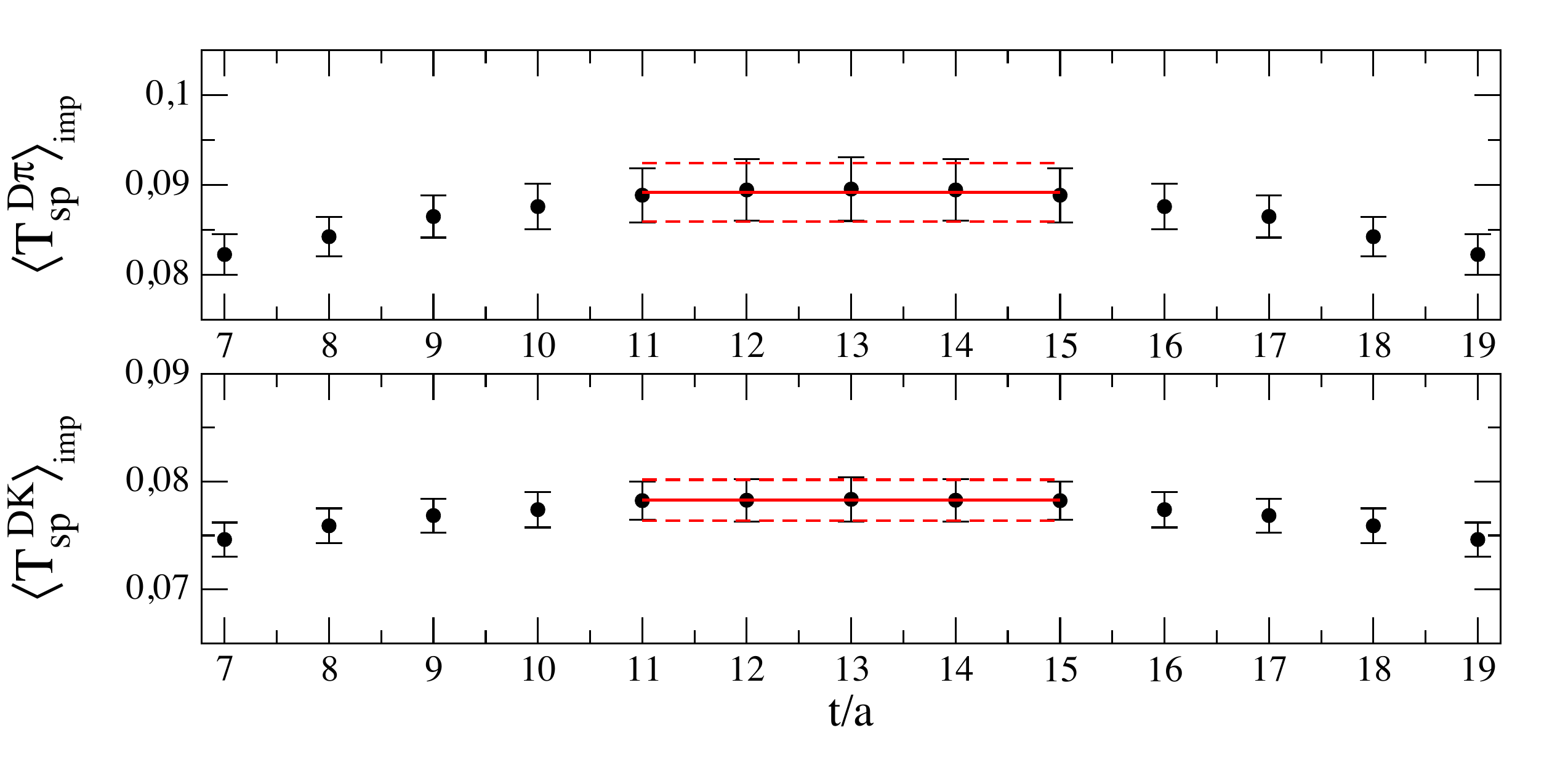}
\vspace{-0.25cm}
\caption{\it \small Matrix elements $\braket{\widehat{T}_{sp}^{D \pi}}_{\rm imp}$ and  $\braket{\widehat{T}_{sp}^{D K}}_{\rm imp}$ extracted from the ratio (\ref{eq:R_plateau}) in the case of the ensemble D20.48 ($\beta = 2.10$, $L / a = 48$) with $a\mu_\ell = 0.0020$, $a\mu_s = 0.0150$, $a\mu_c = 0.170$, $\vec{p}_D = - \vec{p}_{\pi(K)}$ and $|\vec{p}_D| \simeq 150\MeV$. The meson masses are $M_\pi \simeq 255 \MeV$, $M_K \simeq 520 \MeV$  and $M_D \simeq 1640 \MeV$. The solid and dashed red lines correspond to the plateau regions used to extract the matrix elements and to their central values and statistical errors, respectively.}
\label{fig:plateau}
\end{figure}

According to Ref.~ \cite{Frezzotti:2003ni}, in order to guarantee the ${\cal{O}}(a)$ improvement of the extracted matrix elements we consider the two kinematics with opposite spatial momenta of the parent and child mesons and perform the following average:
 \be
     \label{eq:improvement_Ti}
     \braket{\widehat{T}_{sp}^{DP}}_{\rm imp}  \equiv  \frac{1}{2} 
         \left[ \braket{P(E_P, \vec{p}_P) | \widehat{T}_{sp} | D(E_D, \vec{p}_D)} - 
         \braket{P(E_P, -\vec{p}_P) | \widehat{T}_{sp} | D(E_D, -\vec{p}_D)} \right] ~ .
\ee
The quality of the plateau for the matrix elements $\braket{\widehat{T}_{sp}^{D \pi}}_{\rm imp}$ and  $\braket{\widehat{T}_{sp}^{D K}}_{\rm imp}$ is illustrated in Fig.~\ref{fig:plateau}.
The time intervals adopted for fitting Eq.~(\ref{eq:R_plateau}) are symmetric around $t^\prime / 2$ and equal to $[t^\prime / 2 - 2, ~ t^\prime / 2 + 2]$.
These values are compatible with the dominance of the $\pi$, $K$ and $D$ mesons ground-state observed along the time intervals for the two-point correlation functions. To evaluate possible excited states contaminations, we also checked that different choices in the time interval used to fit Eq.~(\ref{eq:R_plateau}), namely $[t^\prime / 2 - 1, ~ t^\prime / 2 + 1]$ and $[t^\prime / 2 - 3, ~ t^\prime / 2 + 3]$, yield changes in the physical tensor form factor that are smaller than the statistical uncertainties.

The standard procedure for determining the tensor form factor $f_T^{DP}(q^2)$ is to assume the Lorentz-covariant decomposition (\ref{eq:matrix_element_decomposition}), which for the current (\ref{eq:Tsp}) implies
 \be
     \label{eq:Tsp_final}
     \braket{\widehat{T}_{sp}^{DP}}_{\rm imp}  = \frac{2}{M_D + M_P} \left[ E_P \braket{p_D} - 
         E_D \braket{p_P} \right] f_T^{DP}(q^2) + {\cal{O}}(a^2) ~ ,
 \ee
where
 \be
     \braket{p_{D(P)}} \equiv \frac{1}{3} \sum_{i = 1}^3 p_{D(P)}^i = \frac{2 \pi}{L} \theta_{D(P)} ~ .
 \ee
In the next Section we present and discuss the result of this determination in which, as anticipated, we find evidence of Lorentz symmetry breaking terms.

%%%%%%%%%%%%%%%%%%%%%%%%%%%%%%%
\section{Tensor form factor and hypercubic effects}
\label{sec:sec2}
%%%%%%%%%%%%%%%%%%%%%%%%%%%%%%%

\begin{figure}[htb!] 
\centering
\makebox[\textwidth][c]{
\includegraphics[width=7.75cm,clip]{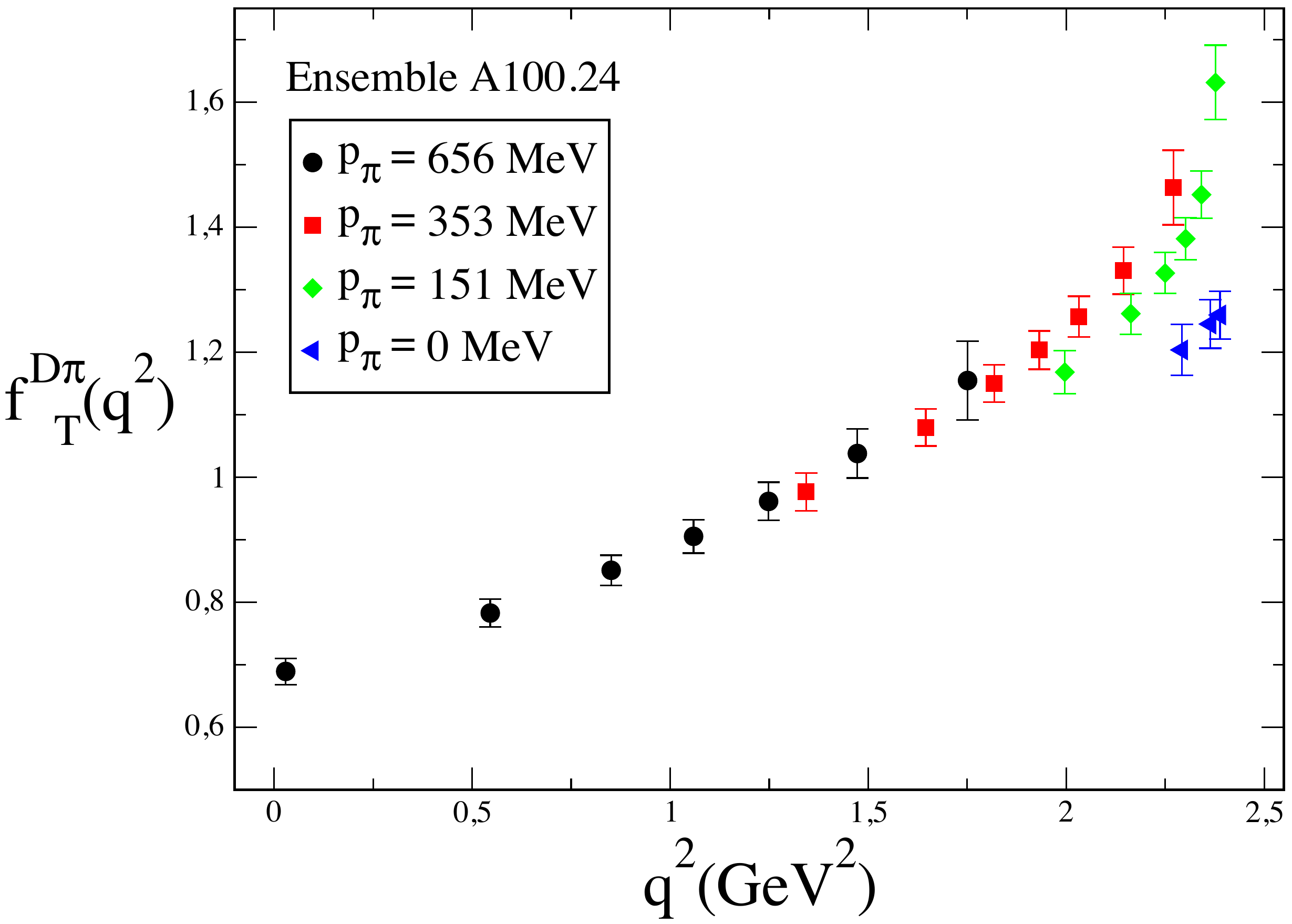}
\includegraphics[width=7.75cm,clip]{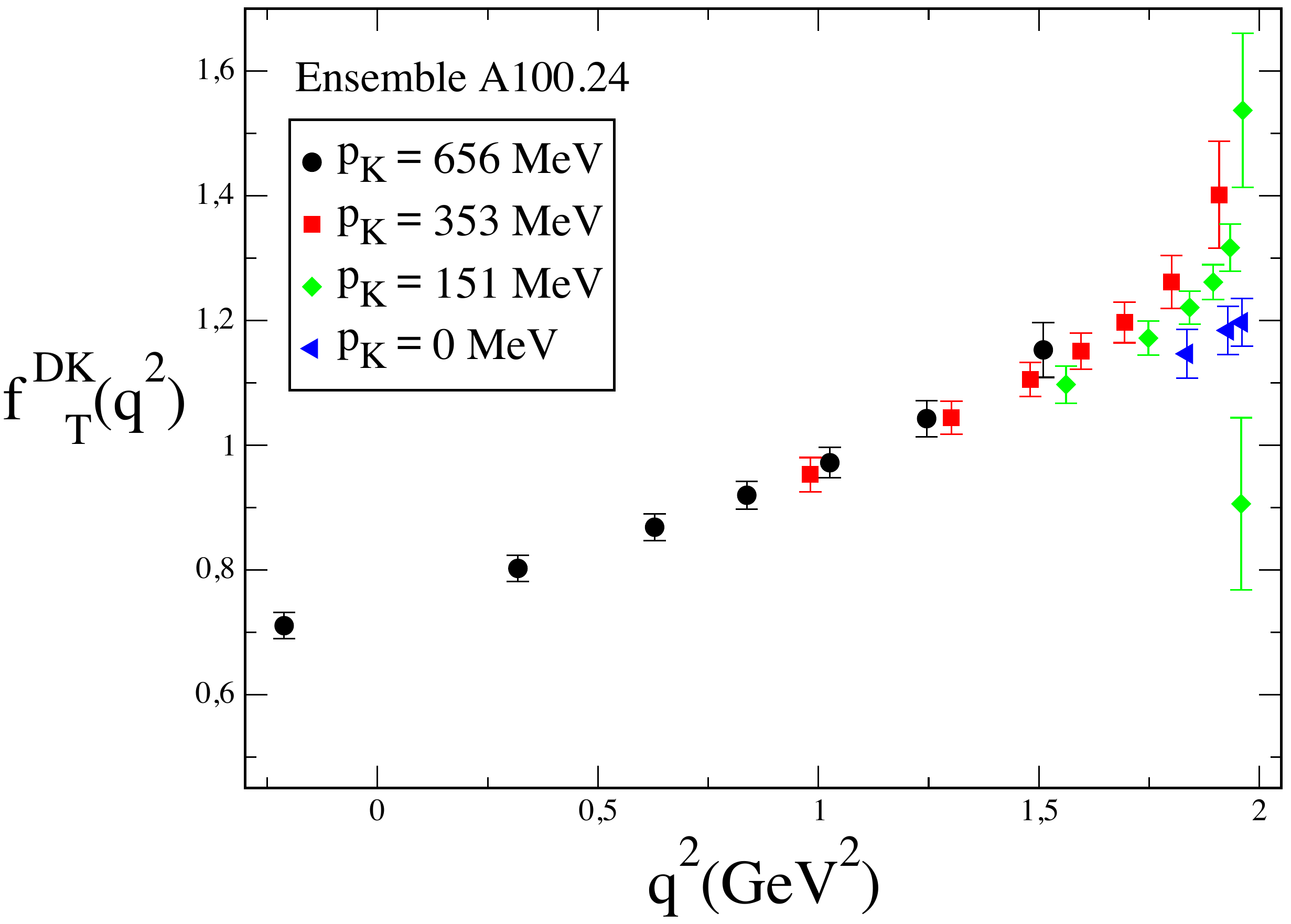}
}
\vspace{-0.75cm}
\caption{\it \small Momentum dependence of the tensor form factors $f_T^{D\pi}$ (left panel) and $f_T^{D K}$ (right panel) in the case of the gauge ensemble A100.24. Different markers and colors distinguish different values of the child meson momentum. The simulated meson masses are $M_\pi \simeq 500$ MeV, $M_K \simeq 639$ MeV and $M_D \simeq 2042$ MeV.}
\label{fig:fishbone}
\end{figure}

After a small interpolation of our lattice data to the physical values of the strange and charm quark masses, $m_s^{phys}(2\,\rm{GeV})=99.6\,(4.3)\,$MeV and $m_c^{phys}(2\,\rm{GeV})=1.176\,(39)\,$GeV taken from Ref.~\cite{Carrasco:2014cwa}, we determine the tensor form factor $f_T^{D \to \pi(K)}(q^2)$ using Eq.~(\ref{eq:Tsp_final}) for each gauge ensemble and for each choice of parent and child meson momenta. 
The momentum dependence of the tensor form factors is illustrated in Fig.~\ref{fig:fishbone}, where different markers and colors correspond to different values of the child meson momentum.
Were the Lorentz-covariant decomposition (\ref{eq:Tsp_final}) adequate to describe the lattice data, the extracted form factor would depend only on the squared four-momentum transfer $q^2$ (and on the parent and child meson masses).
This is not the case and an extra dependence on the value of the child (or parent) meson momentum is clearly visible in Fig.~\ref{fig:fishbone}.
This is due to the fact that the lattice breaks Lorentz symmetry and it is invariant only under discrete rotations by multiple of $90^\circ$ in each direction of the Euclidean space-time. 
Consequently the form factor may depend also on hypercubic invariants. 

In Fig.~\ref{fig:fishbone} finite size effects (FSEs) might in principle play a role. 
However, by comparing the results corresponding to the gauge ensembles A40.24 and A40.32, which share the same pion mass and lattice spacing, but have different lattice sizes ($L = 24 a$ and $L = 32 a$), we found that FSEs are negligible within the current statistical uncertainties.

In Ref.~\cite{Lubicz:2017syv}, where we observed for the first time the breaking of the Lorentz symmetry in the vector and scalar semileptonic $D \to \pi(K)$ form factors, we have proposed a procedure for the subtraction of these effects, which will be applied to the case of the tensor form factor as discussed in the next Section. 

\begin{figure}[htb!]
\centering
\includegraphics[scale=0.40]{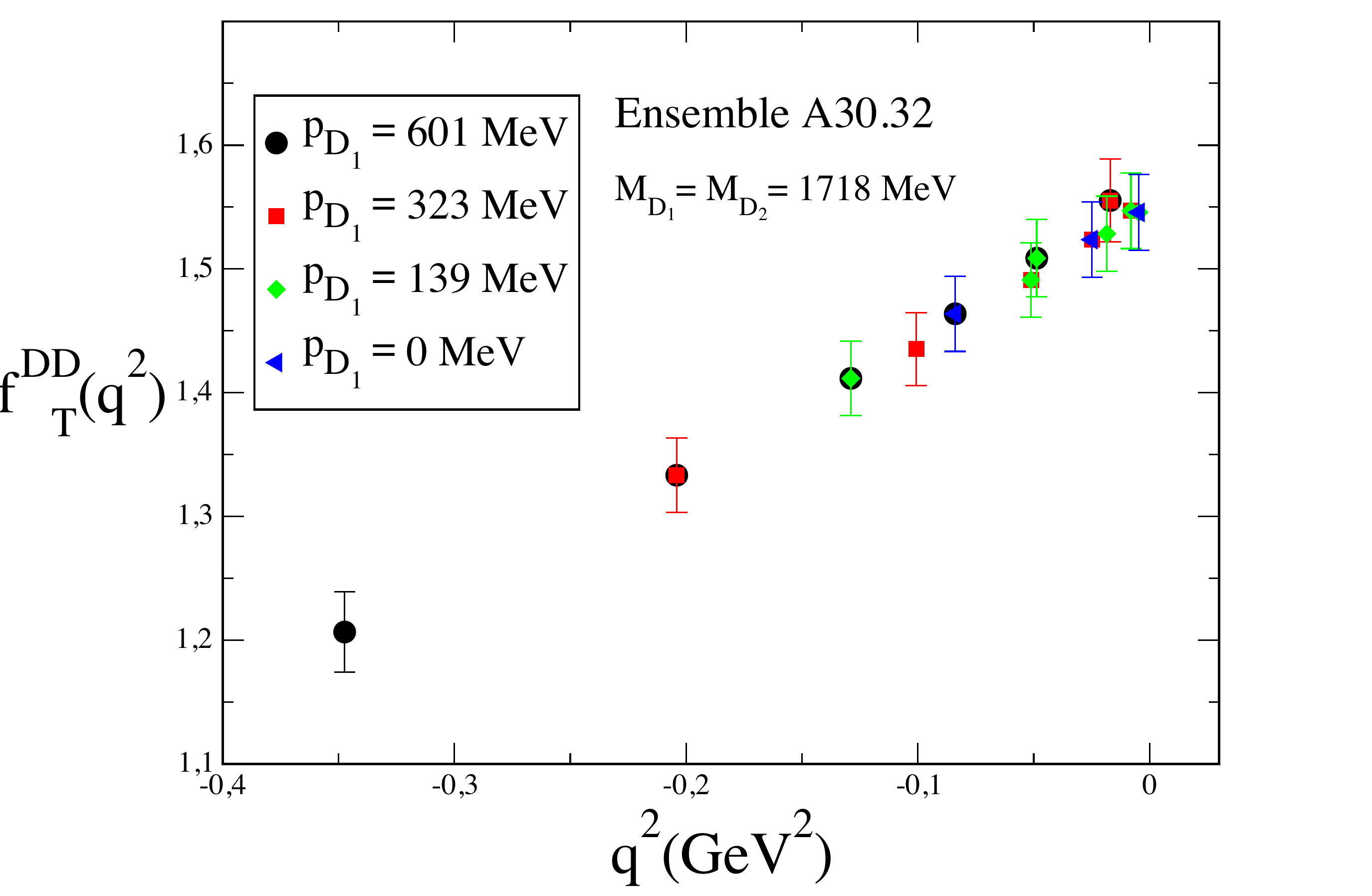}
\vspace{-0.25cm}
\caption{\it \small Momentum dependence of the tensor form factor $f_T^{DD}(q^2)$ in which the parent and child mesons are two charmed P-mesons with degenerate masses equal to $1718$ MeV in the case of the gauge ensemble A30.32. Different markers and colors distinguish different values of the child meson momentum.}
\label{fig:DtoD}
\end{figure}

Before closing this Section, we remind that in Ref~\cite{Lubicz:2017syv} it was observed that the hypercubic artifacts depend largely on the difference between the parent and the child meson masses. 
This finding is confirmed also in the present case of the tensor form factor, as shown in Fig.~\ref{fig:DtoD}.
The momentum dependence of the tensor form factor corresponding to a transition between two mass-degenerate P-mesons shows no evidence of hypercubic effects within the statistical uncertainties.

%%%%%%%%%%%%%%%%%%%%%%%%%%%%%%%
\section{Subtraction of the hypercubic effects}
\label{sec:sec3}
%%%%%%%%%%%%%%%%%%%%%%%%%%%%%%%

Following the approach developed in Ref.~\cite{Lubicz:2017syv} (see there Section 5 for details), we now address the hypercubic effects directly on the matrix elements of the tensor operator. 

We start by introducing Euclidean four-momenta defined, in the case of the four-momentum transfer $q_\mu = (q_0, ~ \vec{q})$, as
 \be
      q_\mu^E = \left( \vec{q}, ~ q_4 \right) = \left( \vec{q}, ~ iq_0 \right) ~ , 
      \label{eq:Euclidean}
 \ee
so that $\sum_\mu q_\mu^E q_\mu^E = - q^2$, and we consider the following decomposition:
 \be
     \braket{P(p_P) | \widehat{T}_{sp}^E | D(p_D)}  =  \braket{P(p_P) | \widehat{T}_{sp}^E | D(p_D)}_{\rm Lor} +  
         \braket{P(p_P) | \widehat{T}_{sp}^E | D(p_D)}_{\rm hyp} ~ ,
    \label{eq:tensor_decomposition}
 \ee
where $\widehat{T}_{sp}^E = i \widehat{T}_{sp}$. 
In the r.h.s.~of Eq.~(\ref{eq:tensor_decomposition}) the first term is the Lorentz-covariant one
 \be
    \braket{P(p_P) | \widehat{T}_{sp}^E | D(p_D)}_{\rm Lor}  = \frac{2}{M_D + M_P} 
        \left[ p_P^4 \braket{p_D} - p_D^4 \braket{p_P} \right] f_T^{DP}(q^2, a^2) ~ ,
    \label{eq:tensor_Lorentz}
 \ee
while $\braket{P(p_P) | \widehat{T}_{sp}^E | D(p_D)}_{\rm hyp}$ describes the hypercubic corrections given by
 \bea
    \braket{P(p_P) | \widehat{T}_{sp}^E | D(p_D)}_{\rm hyp} & = & a^2 \frac{2}{M_D + M_P} 
        \Bigl\{ \left[ (p_P^4)^3 \braket{p_D}  - (p_D^4)^3 \braket{p_P} \right] H_1^{DP}(q^2) \nonumber \\
        & + & \left[ p_P^4 \braket{p_D^{[3]}} - p_D^4 \braket{p_P^{[3]}} \right] H_2^{DP}(q^2) \Bigl\}
    \label{eq:tensor_hypercubic}
 \eea
with $H_1^{DP}(q^2)$ and $H_2^{DP}(q^2)$ being additional hypercubic form factors and $\braket{p_{D(P)}^{[3]}} \equiv (1/3)$ $\sum_{i = 1}^3 (p_{D(P)}^i)^3$.

Eq.~(\ref{eq:tensor_hypercubic}) is the most general structure, up to order ${\cal{O}}(a^2)$, that transforms properly under hypercubic rotations, is antisymmetric under the exchange of the Dirac indices and is built with four powers of the components of the parent and child momenta $p_D^\mu$ and $p_P^\mu$. 
The Lorentz-invariance breaking effects are encoded in the two kinematical structures multiplying the hypercubic form factors $H_j^{DP}$ ($j = 1, 2$).
Consequently, we assume that $H_j^{DP}$ depends only on $q^2$ (and on the parent and child meson masses). 
In particular, we adopt a simple polynomial form
 \be
    H_j(z) = d_0^j + d_1^j z + d_2^j z^2 ~ ,
    \label{eq:Hj}
 \ee
where the $z$ variable is defined as~\cite{Boyd:1995cf,Arnesen:2005ez}
 \be
    z = \frac{{\sqrt {t_ +   - q^2 }  - \sqrt {t_ +   - t_0 } }}{{\sqrt {t_ +   - q^2 }  + \sqrt {t_ +   - t_0 } }}
 \ee
with $t_+$ and $t_0$ given by
 \bea
        t_+  & = & \left( M_D  + M_P \right)^2 ~ , \nonumber \\
        t_0  & = & \left( M_D  + M_P \right) \left( \sqrt {M_D}  - \sqrt {M_P} \right)^2 ~ .
\eea
In Eq.~(\ref{eq:Hj}) the coefficients $d_{0,1,2}^j$ are treated as free parameters and may depend upon the light-quark mass $m_\ell$.

Due to the democratic choice of the spatial components of the meson momenta the hypercubic structure~(\ref{eq:tensor_hypercubic}) cannot be determined and subtracted separately for each gauge ensemble.
Instead it can be fitted by studying simultaneously the data for all the 15 ETMC gauge ensembles used in this work. 
Thus, we performed a global fit considering the dependences on $q^2$, $m_\ell$ and $a^2$ for the tensor form factor $f_T^{DP}$ as well as the $q^2$ and $m_\ell$ dependences of the hypercubic form factors $H_j^{DP}$.

For the form factor $f_T^{DP}(q^2, a^2)$ we adopt the z-expansion of Ref.~\cite{Bourrely:2008za}, modified as in Ref.~\cite{Na:2010uf}, viz.
 \be
   \label{eq:z-exp_fT}
    f_T^{DP}(q^2, a^2) =  \frac{M_D + M_P}{M_D^{\rm phys} + M_P^{\rm phys}}
        \frac{F_T^{DP}(0, a^2) + c_T^{DP}(a^2) (z - z_0) \left(1 + \frac{z + z_0}{2} \right)} 
        {1 - q^2 / \left( M_T^{DP} \right)^2} ~ ,
 \ee
where $z_0 \equiv z(q^2 = 0)$ and the kinematical factor $(M_D + M_P) / (M_D^{\rm phys} + M_P^{\rm phys})$ has been inserted in analogy with the case of the semileptonic $K \to \pi$ transition discussed in Ref.~\cite{Baum:2011rm}~\footnote{We have checked that the results obtained by either including or excluding the kinematical factor $(M_D + M_P) / (M_D^{\rm phys} + M_P^{\rm phys})$ in the r.h.s.~of Eq.~(\ref{eq:z-exp_fT}) are consistent within the statistical errors.}.
In Eq.~(\ref{eq:z-exp_fT}) we consider for the coefficient $c_T^{DP}(a^2)$ a linear dependence on $a^2$, while the pole mass $M_T^{DP}$ is treated as a free parameter in the fitting procedure.
We have checked that making the coefficient $c_T^{DP}$ dependent on the light-quark mass $m_\ell$ does not produce significative changes in the fitting procedure.
Notice that, in the r.h.s.~of Eqs.~(\ref{eq:z-exp_fT}), the term proportional to $z^2$ is not independent of the other terms in the $z$-expansion. 
This constrain comes from the analyticity requirements of $f_T^{DP}$ below the annihilation threshold $\sqrt{t_+} = M_D + M_P$ corresponding to $z = -1$, as described in Ref.~\cite{Bourrely:2008za}. 
We have also explicitly checked that, within the current statistical uncertainties, our data are not sensitive to higher order terms in the z-expansion: the inclusion of a free parameter proportional to $z^2$, plus the appropriate $z^3-$term required by analyticity, produces negligible differences in the final results.
At zero four-momentum transfer, the tensor form factor $f_T^{DP}(0, a^2)$ reduces to
\be
    \label{eq:fT_at_q2_0}
        f_T^{DP}(0, a^2) =  \frac{M_D + M_P}{M_D^{\rm phys} + M_P^{\rm phys}} F_T^{DP}(0, a^2)  ~ .
\ee
For $F_T^{DP}(0, a^2)$ we use the following Ansatz
\be
    \label{eq:ChLim}
    F_T^{DP}(0, a^2) =  F^{DP} \left[ 1 + A^{DP} \xi \log {\xi} + B^{DP} \xi + D^{DP} a^2 \right] ~,
\ee
where $\xi \equiv M_\pi^2 / (4 \pi f_\pi)^2$ and the coefficients $F^{DP}$, $B^{DP}$ and $D^{DP}$ are treated as free parameters in the fitting procedure. 
The chiral-log coefficient $A^{DP}$ is not known in hard pion SU(2) Chiral Perturbation Theory and therefore it is either left as a free parameter or put equal to zero.
The difference between these two Ansatze is used to estimate the systematic uncertainty relative to the chiral extrapolation. 
An extra term proportional to $m_\ell^2$ in Eq.~(\ref{eq:ChLim}) turns out to be consistent with $0$ and, therefore, it is not used for estimating systematic uncertainties.

We have also tried to include an extra term proportional to $(a \Lambda_{\rm{QCD}})^4$. 
Using a value for $\Lambda_{QCD}$ equal to $\simeq 0.35$ GeV, we expect that the value of the coefficient of the extra term is in a natural range of order ${\cal{O}}(1)$. 
Therefore, we adopt for the coefficient of the extra term a (conservative) prior distribution equal to $0 \pm 3$. 
The results obtained using the above alternative fit show marginal differences with respect to the previous ones, and we use it to estimate the systematic uncertainty relative to the discretization effects.  

Using the ingredients described above we fit all the data for the tensor current matrix element $\braket{\widehat{T}_{sp}^{DP}}_{\rm imp}$ coming from the 15 ETMC gauge ensembles and computed at different parent and child momenta. 
The quality of these fits, which include a total of 360 data and a number of free parameters between 9 and 12 depending on the specific Ansatz, is quite good with $\chi^2 / \rm{d.o.f.} \simeq 0.9$ for the $D \to \pi$ transition and $\chi^2 / \rm{d.o.f.} \simeq 0.7$ for the $D \to K$ one. 
The inclusion of the Lorentz breaking terms in the fitting procedure is crucial for obtaining a good description of the data: by putting $H_1^{DP} = H_2^{DP} = 0$ the value of $\chi^2 / \rm{d.o.f.}$ goes up to $\simeq 3.6$ for the $D \to \pi$ case and up to $\simeq 1.8$ for the $D \to K$ transition.

\begin{figure}[htb!] 
\centering
\makebox[\textwidth][c]{
\includegraphics[width=7.75cm,clip]{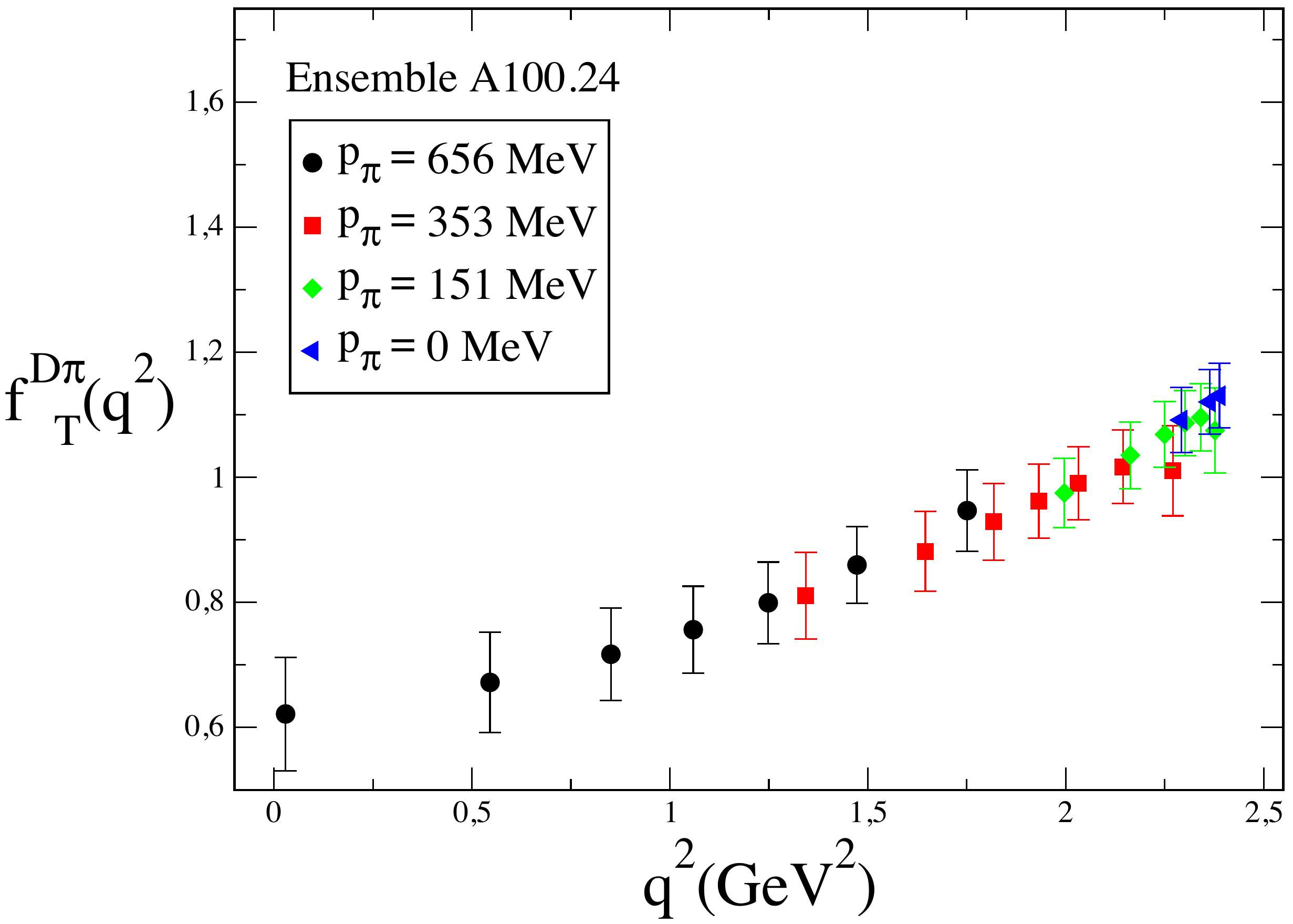}
\includegraphics[width=7.75cm,clip]{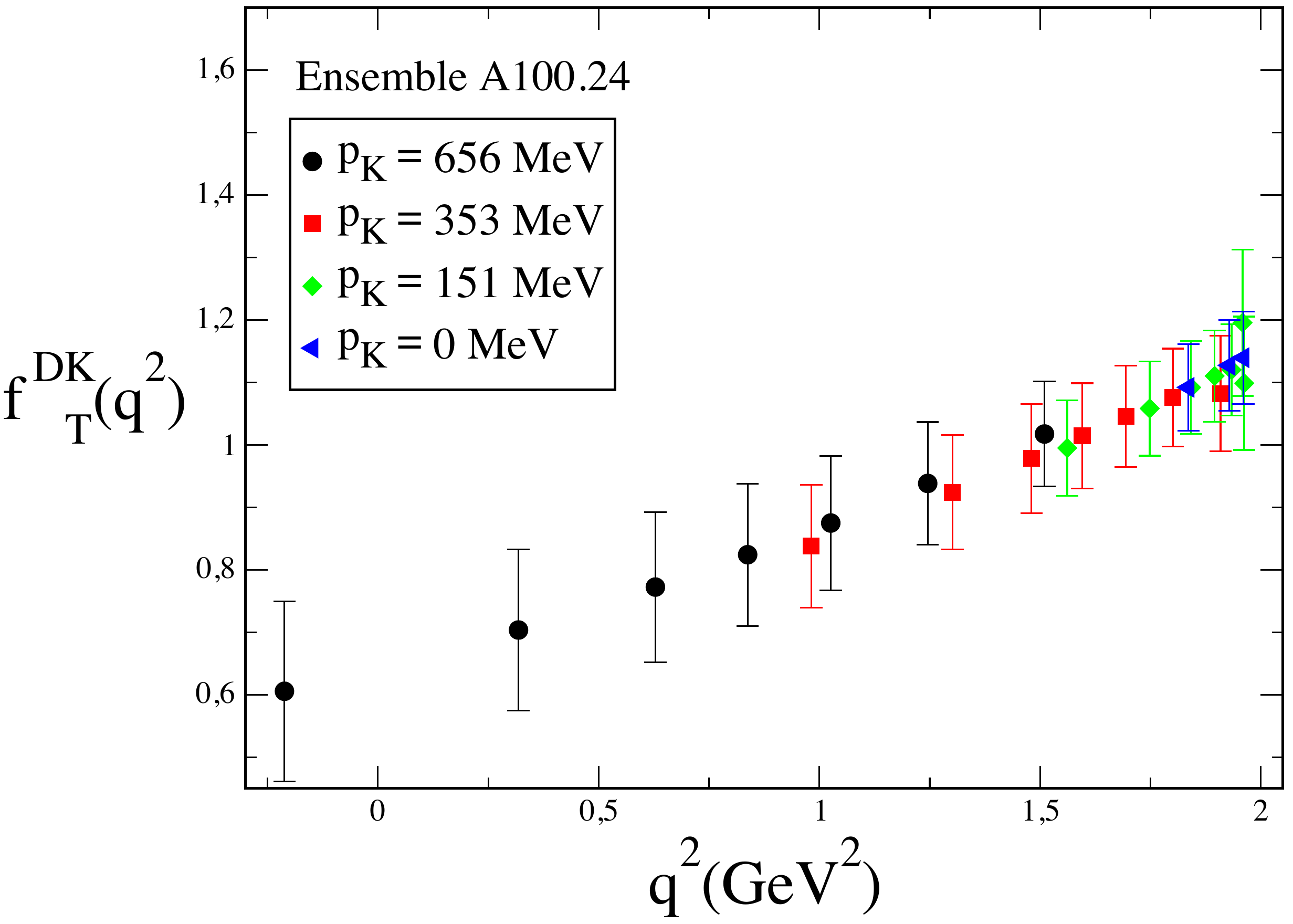}
}
\vspace{-0.75cm}
\caption{\it \small The tensor form factors $f_T^{D\pi}$ (left panel) and $f_T^{DK}$ (right panel) corresponding to the same gauge ensemble of Fig.~\ref{fig:fishbone} after removing the hypercubic effects determined by the global fit.}
\label{fig:corrected}
\end{figure}

In Fig.~\ref{fig:corrected} we show the tensor form factors for the same ensemble used in Fig.~\ref{fig:fishbone} after the subtraction of the hypercubic contributions determined by the global fit. 
It can be seen that the tensor form factors depend now only on the $4-$momentum transfer $q^2$. 

Note that an important feature of our analysis is the use of plenty of kinematical conditions corresponding to parent and child mesons either moving or at rest.
This makes it possible to observe hypercubic effects in the momentum dependence of the data.
As in the case of the vector and scalar form factors~\cite{Lubicz:2017syv}, by considering only a specific reference frame, for instance the Breit-frame in which $\vec{p_D} = - \vec{p}_{\pi(K)}$ or the $D-$meson at rest, the hypercubic effects are present, but not manifest.
\begin{figure}[htb!]
\centering
\makebox[\textwidth][c]{
\includegraphics[width=7.75cm,clip]{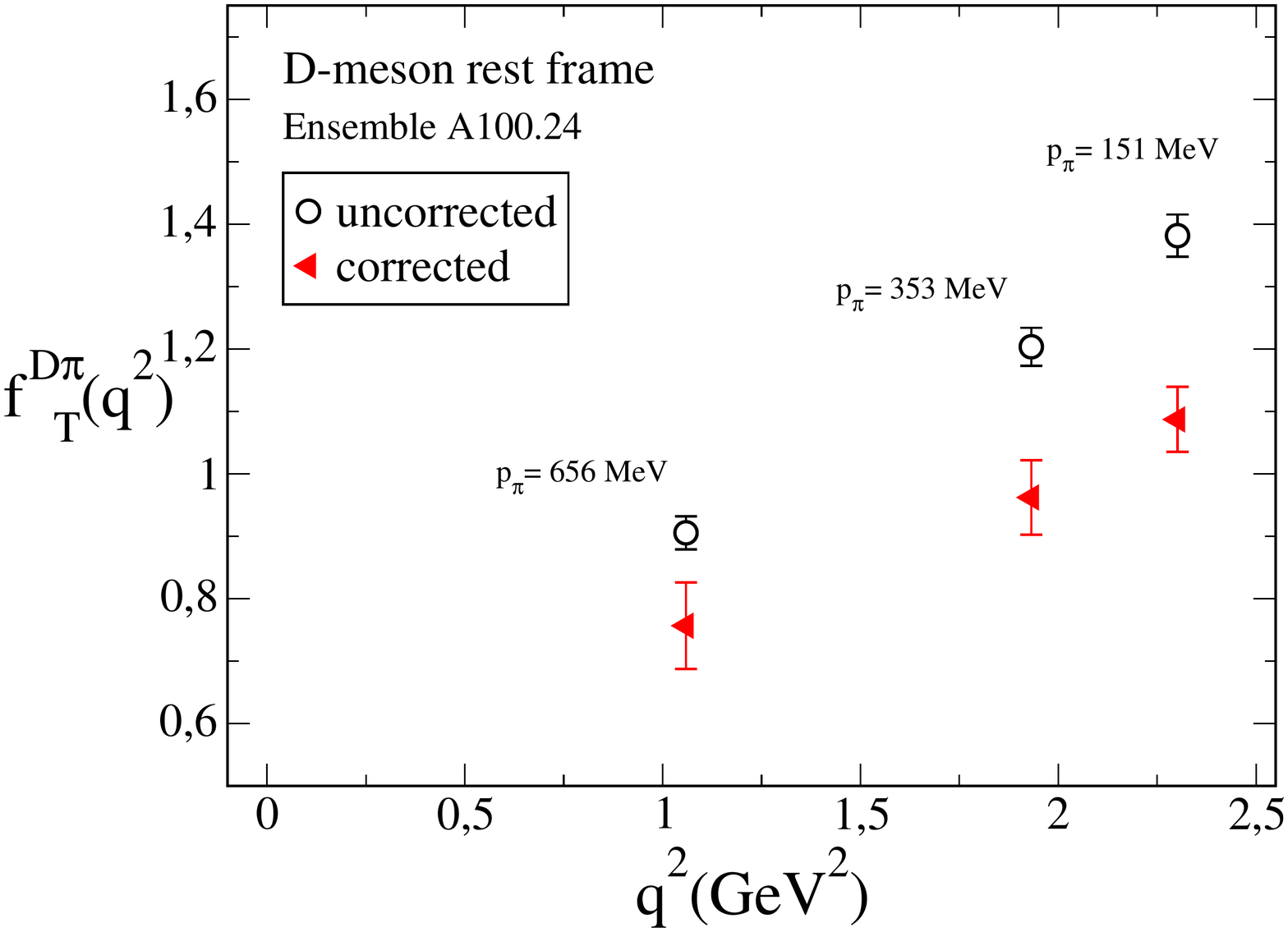}
\includegraphics[width=7.75cm,clip]{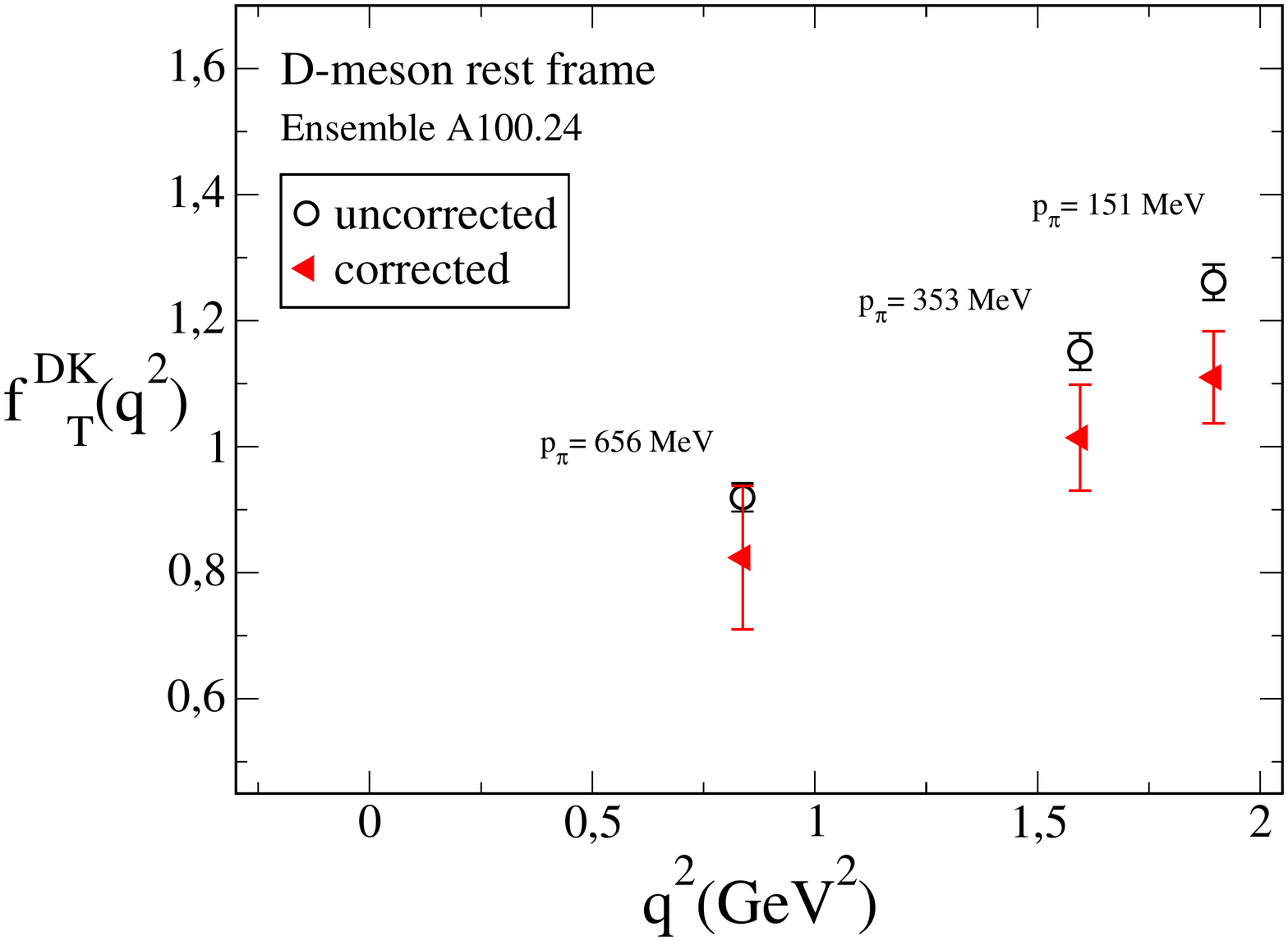}
}
\vspace{-0.75cm}
\caption{\it \small The tensor form factors $f_T^{D \pi}(q^2)$ (left panel) and $f_T^{D K}(q^2)$ (right panel) corresponding to the kinematical conditions with the $D-$meson at rest for the gauge ensemble D30.48. Empty and filled points represent, respectively, the data before and after the removal of the hypercubic effects determined in the global fitting procedure.}
\label{fig:D_at_rest}
\end{figure}
This point is illustrated in Fig.~\ref{fig:D_at_rest}, which shows the subset of our data for the $D \to \pi$ (left panel) and $D\to K$ (right panel) tensor form factors corresponding only to the D-meson at rest both before and after the subtraction of the hypercubic effects determined in the global fitting procedure.
Lorentz-symmetry breaking is not manifest in the limited set of data points with $\vec{p}_D = 0$, but it is not negligible.

In Ref.~\cite{Lubicz:2017syv}, where we studied the hypercubic effects in the vector form factor $f_+(q^2)$, we found that the hypercubic correction is small at $q^2 = 0$ and that the kinematic with the largest correction at high $q^2$ is the one corresponding to the child meson at rest. 
Conversely, in the present case of the tensor form factor $f_T^{DP}(q^2)$, the comparison of the results shown in Figs.~\ref{fig:fishbone} and~\ref{fig:corrected} indicates that mild effects are again present in the low $q^2$ region, while at high $q^2$ the data corresponding to child meson momenta different from zero get the largest hypercubic correction. 
This feature is directly related to the kinematical structures that multiply the hypercubic form factors $H_1^{DP}$ and $H_2^{DP}$ in Eq.~(\ref{eq:tensor_hypercubic}).

%%%%%%%%%%%%%%%%%%%%%%%%%%%%%%%
\section{Results from the global fit}
\label{sec:sec4}
%%%%%%%%%%%%%%%%%%%%%%%%%%%%%%%

\begin{figure}[htb!]
\centering
\includegraphics[scale=0.50]{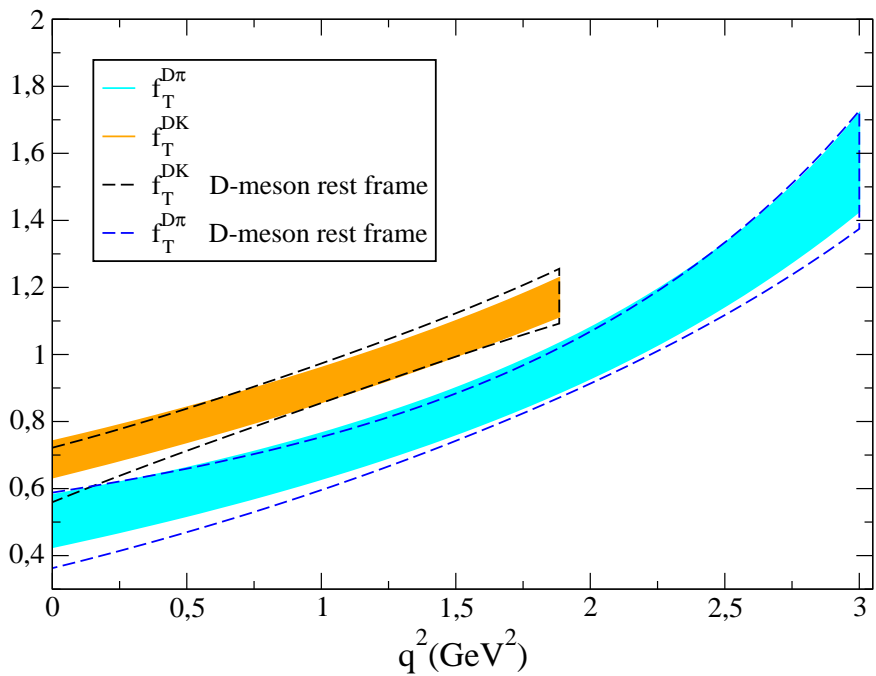}
\vspace{-0.5cm}
\caption{\it \small Comparison of the tensor form factors $f_T^{D\pi}(q^2)$ (cyan solid area) and $f_T^{DK}(q^2)$ (orange solid area), extrapolated to the physical pion mass and to the continuum and infinite volume limits, obtained by choosing all the kinematical configurations and including the hypercubic terms (\ref{eq:tensor_hypercubic}) in the analysis. The dashed lines correspond, instead, to the tensor form factors obtained by limiting to the kinematical configurations corresponding to the $D$-meson rest frame without considering the subtraction of hypercubic effects. All the bands correspond to the total uncertainties at one standard-deviation level.}
\label{fig:comparison_D_at_rest}
\end{figure}

The momentum dependency of the physical Lorentz-invariant tensor form factor, extrapolated to the physical pion mass and to the continuum and infinite volume limits, is shown in Fig.~\ref{fig:comparison_D_at_rest} as a cyan(orange) band for the $D \to \pi(K)$ transition. 

In Fig.~\ref{fig:comparison_D_at_rest} our results for the tensor form factors are compared with those obtained by choosing only the kinematical configurations corresponding to the $D$-meson rest frame and by performing the extrapolations to the physical pion mass and to the continuum and infinite volume limits without including the hypercubic terms (\ref{eq:tensor_hypercubic}). 
In this way, for the limited data set corresponding to the D-meson at rest, the continuum extrapolation is based only on the discretization terms contained in Eq.~(\ref{eq:z-exp_fT}), which are unrelated to hypercubic invariants and correspond simply to $a^2$ or $a^2 q^2$ terms.
This may lead to distortions in the final results at the physical point.
From Fig.~\ref{fig:comparison_D_at_rest} it can be seen that such distortions are found to be comparable with present global uncertainties within one standard-deviation. 
They may become more relevant as the precision of the data will be increased in the future.

In Tables \ref{tab:synthetic_Dpi} and \ref{tab:synthetic_DK} we provide a set of synthetic data points for the tensor $D \to \pi$ and $D \to K$ form factors with the corresponding total uncertainties, calculated for eight selected values of $q^2$ between $0$ and $q^2_{\rm{max}} = (M_D - M_{\pi(K)})^2$. 
We provide also the values of the ratios of the tensor and vector form factors $f_T^{DP}(q^2) / f_+^{DP}(q^2)$ and of the scalar and vector ones $f_0^{DP}(q^2) / f_+^{DP}(q^2)$, using for the latter the results of Ref.~\cite{Lubicz:2017syv}.

\begin{table}[htb!]
\renewcommand{\arraystretch}{1.2} 
{\small
\begin{center}
\begin{tabular}{|c||c|c||c|}
\hline
$q^2 ~ (\mbox{GeV}^2)$ & $f_T^{D\pi}(q^2)$  & $\frac{f_T^{D\pi}(q^2)}{f_+^{D\pi}(q^2)}$ & $\frac{f_0^{D\pi}(q^2)}{f_+^{D\pi}(q^2)}$\\ \hline 
 $0.0~~~$ & ~ $0.506 ~~ (68) ~~ (3) ~ (40) ~ (1) ~~[79]$ & ~ $0.827 ~~ (96) ~ (5) ~ (9) ~ (1) ~ (60)  ~[114]$& $1.000 ~~ [0]$~  \\
 $0.429$ & ~ $0.578 ~~ (62) ~~ (4) ~ (38) ~ (1) ~~[73]$ & ~ $0.807 ~~ (74) ~ (5) ~ (6) ~ (1) ~ (50) ~~[90]$& $0.922 ~~ [8]$~ \\
 $0.857$ & ~ $0.664 ~~ (59) ~~ (5) ~ (36) ~ (1) ~~[69]$ & ~ $0.790 ~~ (58) ~ (4) ~ (4) ~ (1) ~ (41) ~~[71]$& $0.848 ~ [13]$~ \\
 $1.286$ & ~ $0.769 ~~ (59) ~~ (7) ~ (34) ~ (1) ~~[69]$ & ~ $0.775 ~~ (47) ~ (4) ~ (2) ~ (1) ~ (33) ~~[58]$& $0.779 ~ [16]$~  \\
 $1.714$ & ~ $0.899 ~~ (63) ~ (10) ~ (34) ~ (1) ~~[72]$ & ~ $0.763 ~~ (41) ~ (5) ~ (1) ~ (1) ~ (26) ~~[49]$& $0.714 ~ [18]$~ \\
 $2.143$ & ~ $1.065 ~~ (71) ~ (16) ~ (36) ~ (1) ~~[80]$ & ~ $0.752 ~~ (38) ~ (5) ~ (1) ~ (1) ~ (21) ~~[44]$& $0.651 ~ [19]$~  \\ 
 $2.571$ & ~ $1.280 ~~ (85) ~ (27) ~ (43) ~ (1) ~~[95]$ & ~ $0.744 ~~ (41) ~ (6) ~ (3) ~ (1) ~ (18) ~~[45]$& $0.591 ~ [19]$~  \\  
 $3.000$ & ~ $1.573 ~ (122) ~ (46) ~ (58) ~ (1) ~[135]$ & ~ $0.739 ~~ (55) ~ (7) ~ (5) ~ (1) ~ (16) ~~[58]$& $0.533 ~ [18]$~  \\ \hline
\end{tabular}
\end{center}
}
\renewcommand{\arraystretch}{1.0}
\vspace{-0.25cm}
\caption{\it \small Synthetic data points representing our results for the tensor form factor $f_T^{D\pi}(q^2)$ and its ratio with the vector one $f_T^{D\pi}(q^2) / f_+^{D\pi}(q^2)$ (obtained in Ref.~\cite{Lubicz:2017syv}), extrapolated to the physical pion point and to the continuum and infinite volume limits for eight selected values of $q^2$ in the range between $q^2 = 0$ and $q^2 = q_{max}^2 = (M_D - M_\pi)^2 \simeq 3.0 ~ \gev^2$. The errors correspond to the uncertainties related to the statistical + fitting procedure, the input parameters, the chiral extrapolation and the discretization effects, respectively (see text). In the case of the ratio $f_T^{D\pi}(q^2) / f_+^{D\pi}(q^2)$ also the error related to finite size effects (present in the vector form factor~\cite{Lubicz:2017syv}) is shown. The errors in squared brackets correspond to the combination in quadrature of the statistical and all systematic errors. In the rightmost column the synthetic data points corresponding to the ratio of the scalar and vector form factors $f_0^{D\pi}(q^2) / f_+^{D\pi}(q^2)$, as determined in Ref.~\cite{Lubicz:2017syv}, are also shown for comparison.}
\label{tab:synthetic_Dpi}
%\end{table}

\vspace{0.5cm}

%\begin{table}[htb!]
\renewcommand{\arraystretch}{1.2}
{\small
\begin{center}
\begin{tabular}{|c||c|c||c|}
\hline
$q^2 ~ (\mbox{GeV}^2)$ & $f_T^{DK}(q^2)$  & $\frac{f_T^{DK}(q^2)}{f_+^{DK}(q^2)}$ & $\frac{f_0^{DK}(q^2)}{f_+^{DK}(q^2)}$ \\ \hline 
 $0.0~~~$ & ~ $0.687 ~ (51) ~ (15) ~ (10) ~  (1) ~ [54]$ & ~ $0.898 ~ (44) ~ (12) ~ (20) ~  (1) ~ [50]$ & $1.000 ~~ [0]$~  \\
 $0.269$ & ~ $0.741 ~ (50) ~ (15) ~ (10) ~  (1) ~ [53]$ & ~ $0.910 ~ (40) ~ (10) ~ (19) ~  (1) ~ [45]$ & $0.972 ~~ [4]$~  \\
 $0.538$ & ~ $0.799 ~ (48) ~ (15) ~ (10) ~  (1) ~ [52]$ & ~ $0.917 ~ (38) ~~ (7) ~ (17) ~  (1) ~ [42]$ & $0.940 ~~ [7]$~  \\
 $0.808$ & ~ $0.862 ~ (48) ~ (16) ~ (11) ~  (1) ~ [51]$ & ~ $0.920 ~ (36) ~~ (5) ~ (15) ~  (1) ~ [40]$ & $0.906 ~ [10]$~  \\
 $1.077$ & ~ $0.930 ~ (47) ~ (16) ~ (11) ~  (1) ~ [51]$ & ~ $0.918 ~ (34) ~~ (3) ~ (14) ~  (1) ~ [37]$ & $0.868 ~ [13]$~  \\
 $1.346$ & ~ $1.003 ~ (47) ~ (17) ~ (11) ~  (1) ~ [51]$ & ~ $0.911 ~ (31) ~~ (2) ~ (13) ~  (1) ~ [34]$ & $0.827 ~ [16]$~  \\ 
 $1.615$ & ~ $1.083 ~ (48) ~ (19) ~ (11) ~  (1) ~ [53]$ & ~ $0.897 ~ (29) ~~ (1) ~ (12) ~  (1) ~ [31]$ & $0.782 ~ [19]$~  \\  
 $1.885$ & ~ $1.170 ~ (51) ~ (20) ~ (12) ~  (1) ~ [56]$ & ~ $0.876 ~ (28) ~~ (2) ~ (11) ~  (1) ~ [30]$ & $0.733 ~ [21]$~  \\ \hline
\end{tabular}
\end{center}
}
\renewcommand{\arraystretch}{1.0}
\vspace{-0.25cm}
\caption{\it \small The same as in Table~\ref{tab:synthetic_Dpi}, but for the $D \to K$ transition for eight selected values of $q^2$ in the range between $q^2 = 0$ and $q^2 = q_{max}^2 = (M_D - M_K)^2 \simeq 1.88 ~ \gev^2$. }
\label{tab:synthetic_DK}
\end{table}

The errors in Tables \ref{tab:synthetic_Dpi} and \ref{tab:synthetic_DK} take into account the uncertainties induced by:
\begin{itemize}
\item the statistical noise and the fitting procedure itself; we stress that this error coming from a multi-combined fit includes also the uncertainty related to the removal of hypercubic effects;
\item the errors in the determinations of the input parameters of the eight branches of the quark mass analysis of Ref.~\cite{Carrasco:2014cwa}; 
\item the chiral extrapolation, evaluated by combining the results obtained using the SU(2)-inspired Ansatz (\ref{eq:ChLim}) with a free chiral log ($A^{DP} \neq 0$) and without the chiral log ($A^{DP} = 0$); 
\item the (Lorentz-invariant) discretization effects, calculated by comparing the results obtained either including or excluding in Eq.~(\ref{eq:z-exp_fT}) extra terms proportional to $(a \Lambda_{QCD})^4$ and adopting for the value of the corresponding parameters a (conservative) prior distribution equal to $0 \pm 3$.
\end{itemize}

In order to allow a direct use of the synthetic data points without using our bootstrap samples, we have calculated the covariance matrix among the synthetic data points contained either in Table~\ref{tab:synthetic_Dpi} or in Table~\ref{tab:synthetic_DK}.
Moreover, taking into account also the results of Ref.~\cite{Lubicz:2017syv}, we have calculated the full covariance matrix corresponding to the sets of synthetic data points corresponding to all the semileptonic form factors $f_+^{DP}(q^2)$, $f_0^{DP}(q^2)$ and $f_T^{DP}(q^2)$ for $P = \pi$ and $K$ (as well as the full covariance matrix corresponding to our data for $f_+^{DP}(q^2)$, $f_0^{DP}(q^2) / f_+^{DP}(q^2)$ and $f_T^{DP}(q^2) / f_+^{DP}(q^2)$).
The corresponding covariance matrices are available upon request to allow to fit our synthetic data with any functional form, that can be adopted for describing the momentum dependence of the semileptonic form factors.

\begin{figure}[htb!] 
\centering
\makebox[\textwidth][c]{
\includegraphics[width=7.75cm,clip]{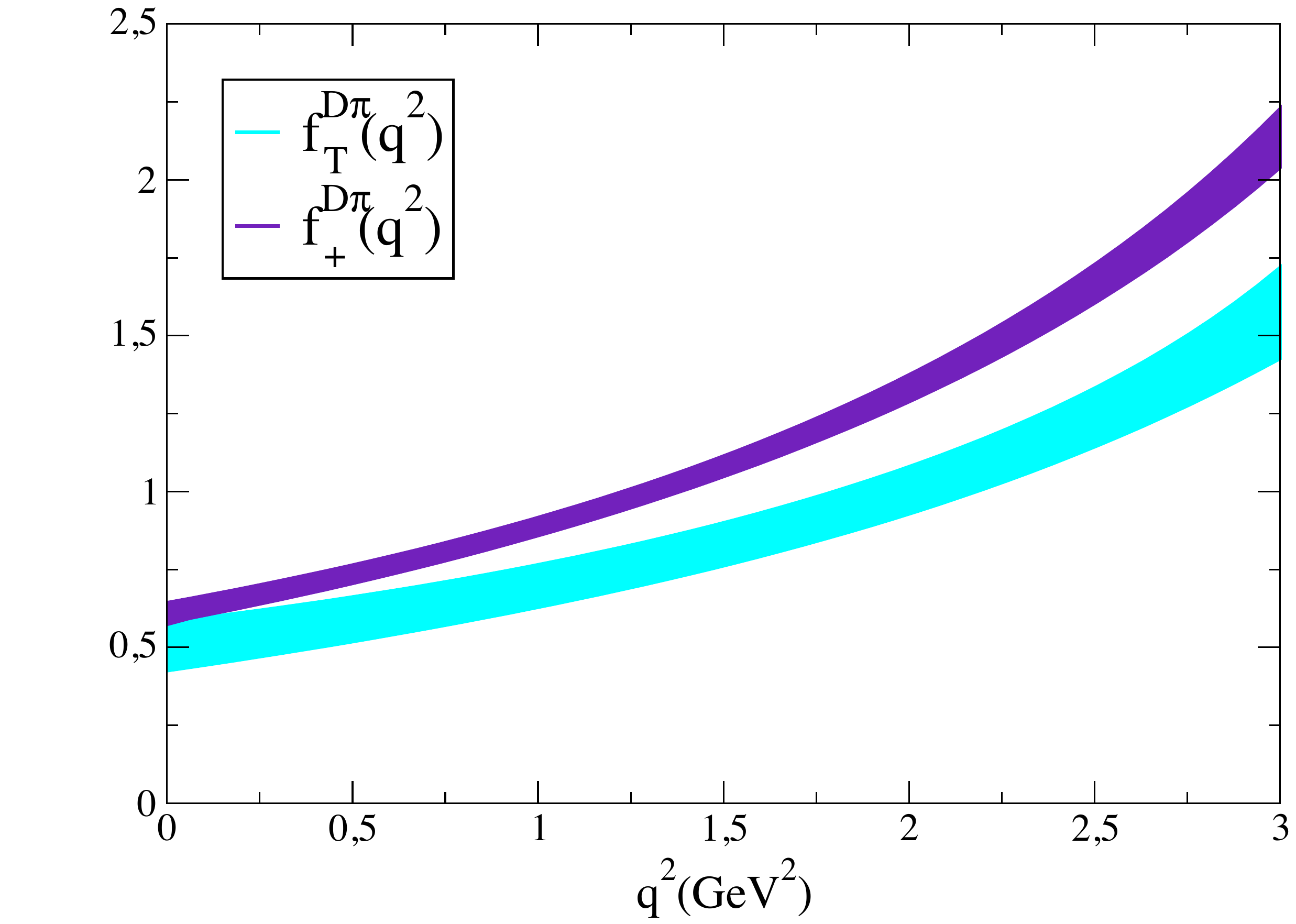}
\includegraphics[width=7.75cm,clip]{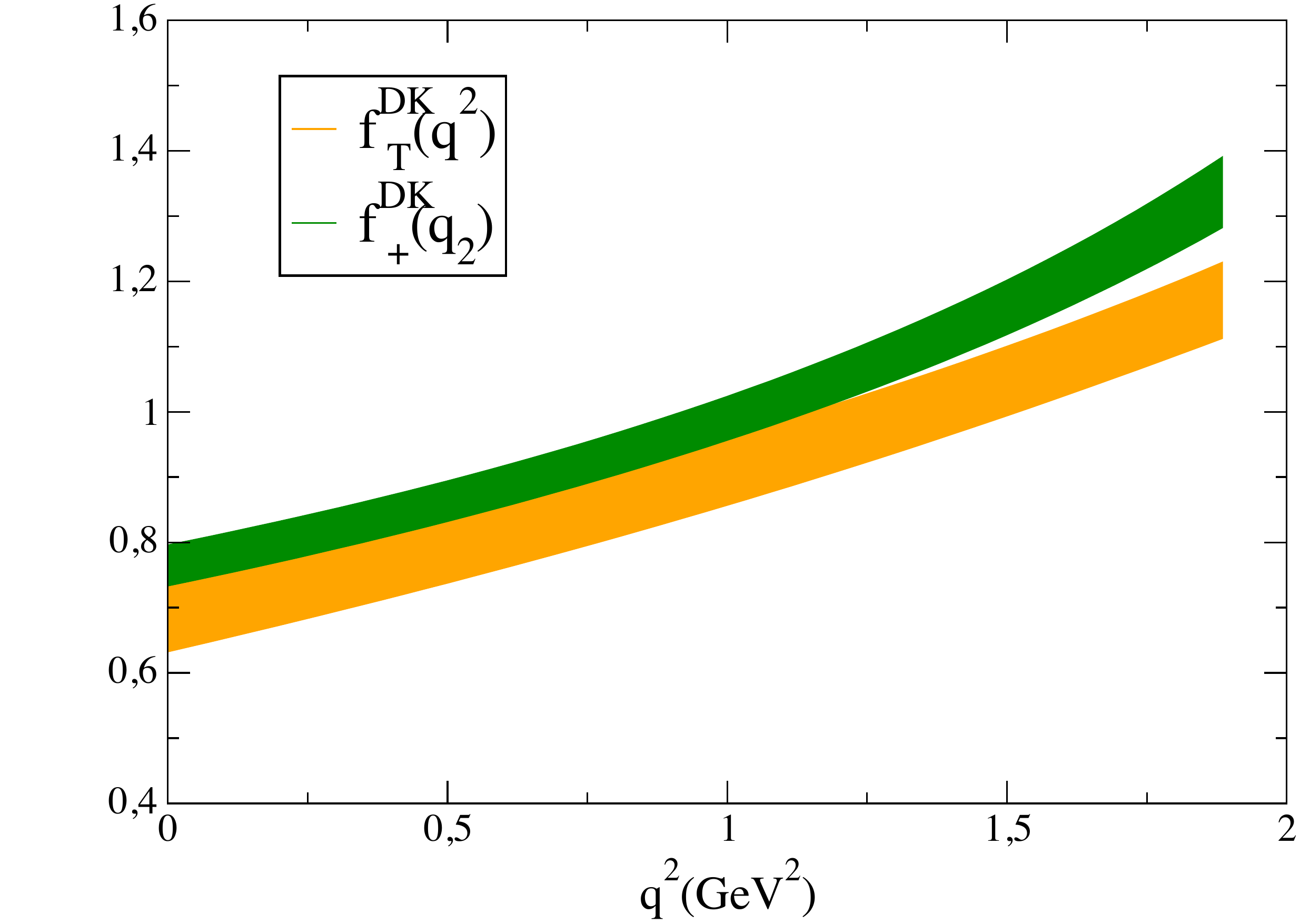}
}
\vspace{-0.75cm}
\caption{\it \small Momentum dependences of the Lorentz-invariant form factor $f_T^{DP}(q^2)$, calculated in this work (cyan bands), and $f_+^{DP}(q^2)$, obtained in Ref.~\cite{Lubicz:2017syv} (orange bands), for the $D \to \pi$ (left panel) and $D \to K$ (right panel) semileptonic transitions. All the form factors are extrapolated to the physical pion mass and to the continuum and infinite volume limits. The bands correspond to the total (statistical + systematic) uncertainty at the level of one standard deviation.}
\label{fig:physicalFormFactors}
\end{figure}

In Fig.~\ref{fig:physicalFormFactors} the tensor form factors $f_T^{D \to \pi(K)}(q^2)$ are compared with the corresponding vector ones $f_+^{D \to \pi(K)}(q^2)$ extracted from the same ETMC gauge ensembles in Ref.~\cite{Lubicz:2017syv}.

%%%%%%%%%%%%%%%%%%%%%%%%%%%%%%%
\section{Conclusions}
\label{sec:conclusions}
%%%%%%%%%%%%%%%%%%%%%%%%%%%%%%%

We have presented the first lattice $N_f = 2 + 1 + 1$ determination of the tensor form factor $f_T^{D \pi(K)}(q^2)$ corresponding to the semileptonic(rare) $D \to \pi(K) \ell \nu_\ell(\ell \ell)$ decays as a function of the squared four-momentum transfer $q^2$.  
Together with the vector $f_+^{D \pi(K)}(q^2)$ and scalar $f_0^{D \pi(K)}(q^2)$ form factors calculated in Ref.~\cite{Lubicz:2017syv}, the present work
completes the set of hadronic matrix elements regulating the semileptonic(rare) $D \to \pi(K) \ell \nu_\ell(\ell \ell)$ transition within and beyond the Standard Model, when a non-zero tensor coupling is possible. 

Our analysis is based on the gauge configurations produced by ETMC with $N_f = 2 + 1 + 1$ flavors of dynamical quarks, which include in the sea, besides two light mass-degenerate quarks, also the strange and charm quarks with masses close to their physical values. 
The matrix elements of the tensor current are determined for a plethora of kinematical conditions in which parent and child mesons are either moving or at rest. 
As in the case of the vector and scalar form factors, Lorentz symmetry breaking due to hypercubic effects is clearly observed also in the data for the tensor form factor and included in the decomposition of the current matrix elements in terms of additional form factors. 

After the extrapolations to the physical pion mass and to the continuum and infinite volume limits we have determined the tensor form factor in the whole kinematical region from $q^2 = 0$ up to $q^2_{\rm max} = (M_D - M_{\pi(K)})^2$ accessible in the experiments.
A set of synthetic data points, representing our results for $f_T^{D \pi(K)}(q^2)$ for several selected values of $q^2$, is provided and the corresponding covariance matrix is also available.
At zero four-momentum transfer our results are
 \be
    \label{eq:form_factors_at_zero}
    f_T^{D\pi}(0) = 0.506 ~ (79) ~ , \qquad \qquad  f_T^{DK}(0) = 0.687 ~ (54)
 \ee
and
  \be
    \label{eq:ratios_at_zero}
    \frac{f_T^{D\pi}(0)}{f_+^{D\pi}(0)} = 0.827 ~ (114) ~ ,  \qquad \qquad \frac{f_T^{DK}(0)}{f_+^{DK}(0)} = 0.898 ~ (50) ~ .
 \ee

\section*{Acknowledgements}
We warmly thank F.~Sanfilippo for his valuable contribution to the initial stage of the present work and our ETMC colleagues for fruitful discussions.
We gratefully acknowledge the CPU time provided by PRACE under the project PRA067 {\it ``First Lattice QCD study of B-physics with four flavors of dynamical quarks"} and by CINECA under the specific initiative INFN-LQCD123 on the BG/Q system Fermi at CINECA (Italy).
V.L., S.S. and C.T.~thank MIUR (Italy) for partial support under Contracts No. PRIN 2010-2011 and No. PRIN 2015.

\appendix
\section*{Appendix: The z-expansion of the tensor form factor at the physical point}

In the case of the $D \to \pi$ transition, after the extrapolations to the physical pion point and to the continuum and infinite volume limits, the z-expansion of the tensor form factor is written as
 \be
   \label{eq:DPi_fT}
    f_T^{D \to \pi}(q^2)  =  \frac{f_T^{D \to \pi}(0) + c_T^{D \to \pi} (z - z_0)
                                             \left(1 + \frac{z + z_0}{2} \right)}{1 - P_T^{D \to \pi} ~ q^2} ~ .
 \ee
The values of the three parameters $f_T^{D \to \pi}(0)$, $c_T^{D \to \pi}$, $P_T^{D \to \pi}$, are collected in Table~\ref{tab:Dpi_parms}, with the corresponding covariance matrix given in Table~\ref{tab:Dpi_cov}.

\begin{table}[htb!]
\renewcommand{\arraystretch}{1.2} 
\begin{center}
\begin{tabular}{|c|c|c|}
\hline
$f^{D \to \pi}(0)$ & $c_T^{D \to \pi}$ & $P_T^{D \to \pi}~(\mbox{GeV}^{-2})$   \\ \hline 
$0.5063 ~ (786)$ & $-1.10 ~ (1.03)$ & $0.1461 ~ (681)$  \\
\hline
\end{tabular}
\end{center}
\renewcommand{\arraystretch}{1.0}
\vspace{-0.25cm}
\caption{\it \small Values of the parameters appearing in the z-expansion of the tensor form factors (\ref{eq:DPi_fT}) in the case of the $D \to \pi$ transition.}
\label{tab:Dpi_parms}
\end{table}

\begin{table}[htb!]
\small
\renewcommand{\arraystretch}{1.2} 
\begin{center}
\begin{tabular}{|c||c|c|c|}
\hline
                               & $f_T^{D \to \pi}(0)$       & $c_T^{D \to \pi}$         & $P_T^{D \to \pi}$   \\ \hline 
$f_T^{D \to \pi}(0)$ & $  6.183 \cdot 10^{-3}$ & $3.995 \cdot 10^{-2}$ & $1.472 \cdot 10^{-3}$ \\ 
$c_T^{D \to \pi}$    & $  3.995 \cdot 10^{-2}$ & $1.059   $                    & $6.637 \cdot 10^{-2}$ \\ 
$P_T^{D \to \pi}$   & $  1.472 \cdot 10^{-3}$ & $6.637 \cdot 10^{-2}$  & $4.632 \cdot 10^{-3}$ \\ 
\hline
\end{tabular}
\end{center}
\renewcommand{\arraystretch}{1.0}
\vspace{-0.25cm}
\caption{\it \small Covariance matrix corresponding to the z-expansions of the tensor form factor (\ref{eq:DPi_fT}) in the case of the $D \to \pi$ transition.}
\label{tab:Dpi_cov}
\end{table}

Analogously, in the case of the $D \to K$ transition the z-expansion of the tensor form factor reads as 
 \be
   \label{eq:DK_fT}
    f_T^{D \to K}(q^2)  =  \frac{f_T^{D \to K}(0) + c_T^{D \to K} (z - z_0)
                                          \left(1 + \frac{z + z_0}{2} \right)}{1 - P_T^{D \to K} ~ q^2} ~ ,
 \ee
where the values of the three parameters $f_T^{D \to K}(0)$, $c_T^{D \to K}$ and $P_T^{D \to K}$ are collected in Table~\ref{tab:DK_parms}, with the corresponding covariance matrix given in Table~\ref{tab:DK_cov}.

\begin{table}[htb!]
\renewcommand{\arraystretch}{1.2} 
\begin{center}
\begin{tabular}{|c|c|c|}
\hline
$f^{D \to K}(0)$ & $c_T^{D \to K}$ & $P_T^{D \to K}~(\mbox{GeV}^{-2})$   \\ \hline 
$0.6871 ~ (542)$ & $-2.86 ~ (1.46)$ & $0.0854 ~ (671)$  \\
\hline
\end{tabular}
\end{center}
\renewcommand{\arraystretch}{1.0}
\vspace{-0.25cm}
\caption{\it \small Values of the parameters appearing in the z-expansion of the tensor form factor (\ref{eq:DK_fT}) in the case of the $D \to K$ transition.}
\label{tab:DK_parms}
\end{table}

\begin{table}[htb!]
\renewcommand{\arraystretch}{1.2} 
\begin{center}
\begin{tabular}{|c||c|c|c|}
\hline
                   & $f^{D \to K}(0)$        & $c_T^{D \to K}$         & $P_T^{D \to K}$ \\ \hline 
$f^{D \to K}(0)$   & $2.938  \cdot 10^{-3}$  & $9.605 \cdot 10^{-3}$   & $-2.231 \cdot 10^{-4}$ \\ 
$c_+^{D \to K}$    & $9.605  \cdot 10^{-3}$  & $2.125              $   & $ 9.298 \cdot 10^{-2}$ \\ 
$P_T^{D \to K}$    & $-2.231 \cdot 10^{-4}$  & $9.298 \cdot 10^{-2}$   & $ 4.505 \cdot 10^{-3}$ \\ 
 \hline
\end{tabular}
\end{center}
\renewcommand{\arraystretch}{1.0}
\vspace{-0.25cm}
\caption{\it Covariance matrix corresponding to the z-expansions of the tensor form factor (\ref{eq:DK_fT}) in the case of the $D \to K$ transition.}
\label{tab:DK_cov}
\end{table}

\bibliographystyle{JHEP}

\bibliography{rifbiblio}

\end{document}